\documentclass[aps,amsfonts,amsmath,amssymb,nofootinbib,twocolumn,superscriptaddress]{revtex4}

\usepackage{graphics}
\usepackage{hyperref}
\usepackage{epsfig}
\usepackage{color}
\usepackage{physics}

\usepackage[normalem]{ulem}
\usepackage{multirow}
\usepackage{bm}
\usepackage{soul}
\usepackage{titlesec}
\usepackage{hyperref}

\hypersetup{draft}

\begin{document}

\bibliographystyle{naturemag}

\title{Imaging the N\'eel vector switching in the monolayer antiferromagnet MnPSe$_3$ with strain-controlled Ising order} 

\author{Zhuoliang Ni}
\affiliation{Department of Physics and Astronomy, University of Pennsylvania, Philadelphia, Pennsylvania 19104, U.S.A}
\author{A. V. Haglund}
\affiliation{Department of Materials Science and Engineering, University of Tennessee, Knoxville, TN 37996, U.S.A.}
\author{H. Wang}
\affiliation{Department of Materials Science and Engineering, Texas A\&M University, College Station, TX 77843 U.S.A}
\author{B. Xu}
\affiliation{Department of Physics and Fribourg Center for Nanomaterials, University of Fribourg, Chemin du Mus\'{e}e 3, CH-1700 Fribourg, Switzerland}
\author{C. Bernhard}
\affiliation{Department of Physics and Fribourg Center for Nanomaterials, University of Fribourg, Chemin du Mus\'{e}e 3, CH-1700 Fribourg, Switzerland}
\author{D. G. Mandrus}
\affiliation{Department of Materials Science and Engineering, University of Tennessee, Knoxville, TN 37996, U.S.A.}
\affiliation{Materials Science and Technology Division, Oak Ridge National Laboratory, Oak Ridge, TN, 37831, U.S.A.}
\author{X. Qian}
\affiliation{Department of Materials Science and Engineering, Texas A\&M University, College Station, TX 77843 U.S.A}
\author{E. J. Mele}
\affiliation{Department of Physics and Astronomy, University of Pennsylvania, Philadelphia, Pennsylvania 19104, U.S.A}
\author{C. L. Kane}
\affiliation{Department of Physics and Astronomy, University of Pennsylvania, Philadelphia, Pennsylvania 19104, U.S.A}
\author{Liang Wu}
\email{liangwu@sas.upenn.edu}
\affiliation{Department of Physics and Astronomy, University of Pennsylvania, Philadelphia, Pennsylvania 19104, U.S.A}

\date{\today}

\maketitle

\textbf{The family of monolayer two-dimensional (2D) materials hosts a wide range of interesting phenomena, including superconductivity \cite{costanzonatnano16}, charge density waves \cite{xinatnano15}, topological states \cite{wuscience18} and ferromagnetism \cite{huangnat17}, but direct evidence for antiferromagnetism in the monolayer has been lacking \cite{maknatrevphys19}. Nevertheless, antiferromagnets have attracted enormous interest recently in spintronics due to the absence of stray fields and their terahertz resonant frequency \cite{nvemecnatphys18}. Despite the great advantages of antiferromagnetic spintronics,  controlling and  detecting  N\'eel vectors have been limited in bulk materials \cite{wadleyscience16, saidlnatphot17, nvemecnatphys18, cheongnpj20, nairnatmat20}. In this work, we developed a sensitive second harmonic generation  (SHG) microscope and detected long-range N\'eel antiferromagnetic (AFM) order and N\'eel vector switching down to the  monolayer in MnPSe$_3$.    Temperature-dependent SHG measurement in repetitive thermal cooling surprisingly collapses into two curves, which correspond to the switching of an Ising type  N\'eel vector  reversed by the time-reversal operation,  instead of a six-state clock ground state expected from the threefold rotation symmetry in the structure \cite{oshikawaprb00, louprl07, cheongnpj19}.  We imaged the spatial distribution of the  N\'eel vectors across samples and rotated them by an arbitrary angle irrespective of the lattice in the sample plane by  applying strain. By studying both a Landau theory and a microscopic model that couples strain to nearest-neighbor exchange, we conclude that the phase transition of the XY model in the presence of strain falls into the Ising universality class instead of the XY one, which could explain the extreme strain tunability. Finally, we found that the 180$^\circ$ AFM domain walls  are highly mobile down to the monolayer after thermal cycles, paving the way for future control of the antiferromagnetic domains by strain or external fields on demand for ultra-compact  2D AFM terahertz spintronics.}

Detection and control of the spin order in ferromagnetic materials is the main principle in current information technology. The discovery of  2D ferromagnetic materials using the polar Kerr effect \cite{huangnat17,gongnat17}  has triggered tremendous interest in studying magnetism in the true 2D limit \cite{dengnat18, feinatmat18, thielsci19, chensci19}  and spintronic device applications in van der Waals heterostructure materials \cite{songsci18,kleinsci18,wangnatcomm18, huangnatnano18, jiangnatmat18, jiangnatnano18}.  Optical techniques are powerful tools to detect magnetism \cite{maknatrevphys19}, but  clear evidence for the AFM order in atomically thin 2D crystals has not been identified due to the lack of  sensitive direct detection.  The polar Kerr effect is absent in AFM materials when the total magnetization is zero \cite{burchnat18, gongsci19, maknatrevphys19, gibertininatnano19}. Although Raman spectroscopy is a powerful tool to study spin-phonon coupling and collective magnons \cite{leenanolett16, wang2Dmat16, kim2dmat19, vaclavkova2dmat20}, their identification often do not provide  unambiguous identification  of the AFM order \cite{maknatrevphys19, huangnatmat2020}. Non-optical methods such as tunneling magneto-resistance measurement have indicated the correlation in the monolayer of an AFM material\cite{longnanoletter20}, but it is also not a direct probe of the AFM order parameter \cite{maknatrevphys19,huangnatmat2020}. SHG has been shown to be a sensitive tool to detect AFM orders due to inversion symmetry breaking from the spin order in magneto-electric  materials including bulk Cr$_2$O$_3$ \cite{fiebigreview05}, few-layer MnPS$_3$ \cite{chuprl20} and a synthetic bilayer CrI$_3$ \cite{sunnat19}.  Nevertheless, the detection of intrinsic AFM in the monolayer has  not been demonstrated yet, which remains an unresolved fundamental question and is also not ideal for AFM terahertz spintronics at the smallest scales. In this work, we  systematically study the layer-dependent AFM order, N\'eel vector distribution, switching and its strain tunability in a 2D crystal of MnPSe$_3$ \cite{wiedenmannssc81} by a newly developed sensitive SHG imaging microscope. (See methods.)

\begin{figure*}
\centering
\includegraphics[width=\textwidth]{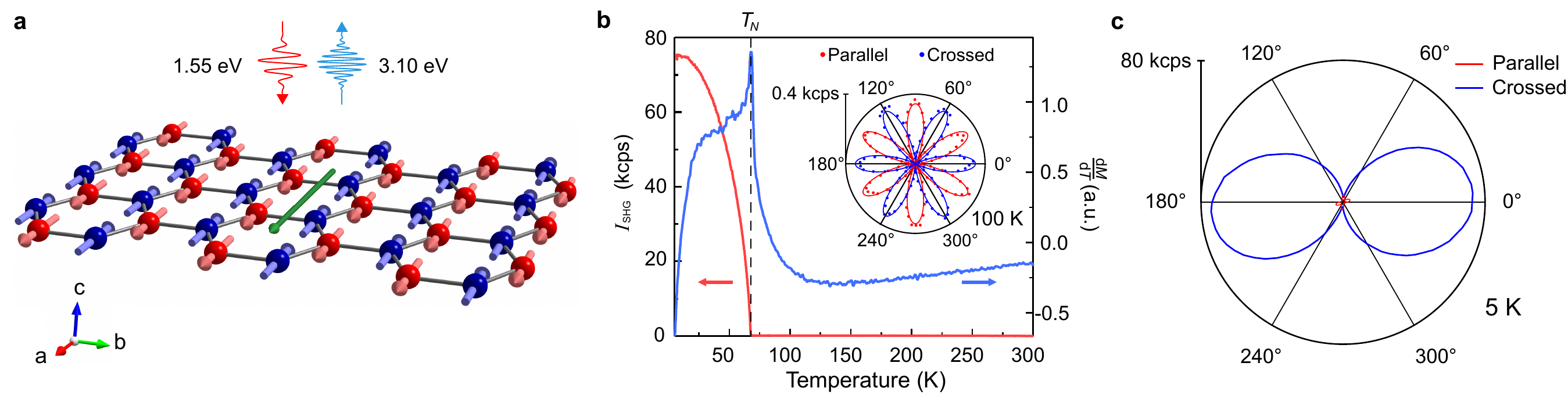}
\caption{\textbf{Characterization of bulk MnPSe$_3$ samples.} {\bf a}, A schematic of the in-plane AFM N\'eel order and SHG measurement on MnPSe$_3$. The Mn atoms in each layer form a honeycomb lattice. The green arrow denotes the N\'eel vector. Laser pulses centered at 1.55 eV are normally incident on the $ab$ plane and reflected light at the second-harmonic frequency is detected.  {\bf b}, Temperature dependence of SHG intensity (red) on a 100-nm sample exfoliated on SiO$_2$/Si and the temperature derivative of the in-plane magnetic susceptibility of a bulk crystal (blue). `kcps' stands for one thousand counts per second.  The transition temperature is marked by the black dashed line. Inset: Polarization-resolved EQ SHG of the 100-nm flake at 100 K. The dots are the data and the solid lines are the fits. {\bf c}, Polarization-resolved SHG polar patterns of the same 100-nm thick MnPSe$_3$ flake measured at 5 K  (All of the lines are experimental data). $0^{\circ}$ is defined as the angle between the polarization of the incidence pulse and the horizontal direction in the lab.}
\label{Fig1}
\end{figure*}

MnPSe$_3$ belongs to the family of AFM transition metal phosphorous trichalcogenide MPX$_3$ (M = Mn, Ni, Fe, Co, X = S, Se), among which Mn compounds form inversion-breaking N\'eel order while others are in centrosymmetric Zigzag ordered phases \cite{wiedenmannssc81, wildesjpcm98, wildesjpcm12, wildesprb15, ressoucheprb10, lancconprb16}. In contrast to MnPS$_3$, which has dominant out-of-plane moments \cite{jeevanandamjpcm1999,ressoucheprb10}, MnPSe$_3$ has in-plane spins with very large XY anisotropy according to the neutron scattering measurement \cite{wiedenmannssc81} (Fig. ~\ref{Fig1}\textbf{a}), which offers richer magnetic domain structures such as vortices and tunability \cite{cheongnpj19}. Above the N\'eel temperature, $T_N$, MnPSe$_3$ belongs to the point group $\overline{3}$ $(S_6)$ and space group 148 \cite{wiedenmannssc81}, and has an inversion center between two neighboring Mn atoms but no mirror symmetry.  The Mn atoms form a honeycomb lattice in one layer (Fig. ~\ref{Fig1}\textbf{a}), and the honeycomb layers form the rhombohedral (ABC) stacking along the $c$ axis with a threefold rotational symmetry. Different from FePS$_3$ and NiPS$_3$, which have a change of twofold rotational symmetry in the bulk to a threefold rotational symmetry in the monolayer \cite{leenanolett16, wang2Dmat16, kimnatcomm19, kangnat2020}, MnPSe$_3$ is always three-fold symmetric. As shown in the inset of Fig. ~\ref{Fig1}\textbf{b}, a small temperature-independent SHG  from the electric quadruple (EQ) contribution, $I^{EQ}_i(2\omega) \sim | \sum_{jkl} \chi^{EQ}_{ijkl}E_j(\omega)\nabla_kE_l(\omega) |^2$ follows the lattice threefold rotational symmetry above $T_N$. (See methods.) The parallel and crossed configuration correspond to $E(2\omega) // E(\omega)$ and $E(2\omega) \perp E(\omega)$ respectively while we co-rotate $E(\omega)$ and $E(2\omega)$ by 360$^{\circ}$ in the $ab$ plane \cite{wunatphys17}.  Below the N\'eel temperature, the formation of the N\'eel AFM order with in-plane spins breaks the inversion symmetry ($\mathcal{P}$), which allows electric dipole (ED) contribution to the SHG, $I^{ED}_i(2\omega) \sim | \sum_{jk} \chi^{ED}_{ijk}(\mathbf{L}) E_j (\omega) E_k (\omega) |^2$ \cite{fiebigreview05}. $\chi^{ED}_{ijk}(\mathbf{L})$ is proportional to the order parameter, the N\'eel vector $\mathbf{L}$, \cite{savalenti00} and changes the sign when $\mathbf{L}$ flips by 180$^{\circ}$ ($\mathbf{L}=\mathbf{M_1}-\mathbf{M_2}$, where $\mathbf{M_1}$ and $\mathbf{M_2}$ are the magnetization of two neighboring Mn atoms). (See Fig. ~\ref{Fig1}\textbf{a}). In the AFM state,  the product of the inversion symmetry  ($\mathcal{P}$) and the time-reversal symmetry ($\mathcal{T}$), so called $\mathcal{PT}$ symmetry,  is still preserved \cite{lipnas13},  even though both $\mathcal{P}$ and $\mathcal{T}$ are broken. This kind of ED term is often called non-reciprocal or c-type SHG allowed by the $\mathcal{PT}$ symmetry, while the EQ term is an i-type SHG, where `c' and `i' mean changing  and invariant under time-reversal symmetry respectively \cite{fiebigreview05}. Fig. \ref{Fig1}\textbf{b} shows a typical SHG response as a function of temperature on a $\sim$ 100-nm thick flake exfoliated on SiO$_2$/Si. A sharp turn-on of the ED SHG signal clearly indicates a phase transition at 67.9 $\pm$ 0.2 K, agreeing well with the $T_N$ (68 $\pm$ 0.5 K) determined from in-plane magnetization measurement. Below $T_N$, as shown in Fig. \ref{Fig1}\textbf{c}, a giant twofold signal emerges in the crossed configuration, which clearly breaks the threefold rotation symmetry.  Another surprising observation is that the peak of the parallel polar pattern is only 1/20 of that in the crossed pattern, which was not observed in previous $\mathcal{PT}$  invariant van der Waals AFM materials \cite{sunnat19, chuprl20}.   The nodal direction of the twofold crossed pattern is also shown to be close to the N\'eel vector direction (see Supplementary Note 1 and 2).  

\begin{figure*}
\centering
\includegraphics[width=\textwidth]{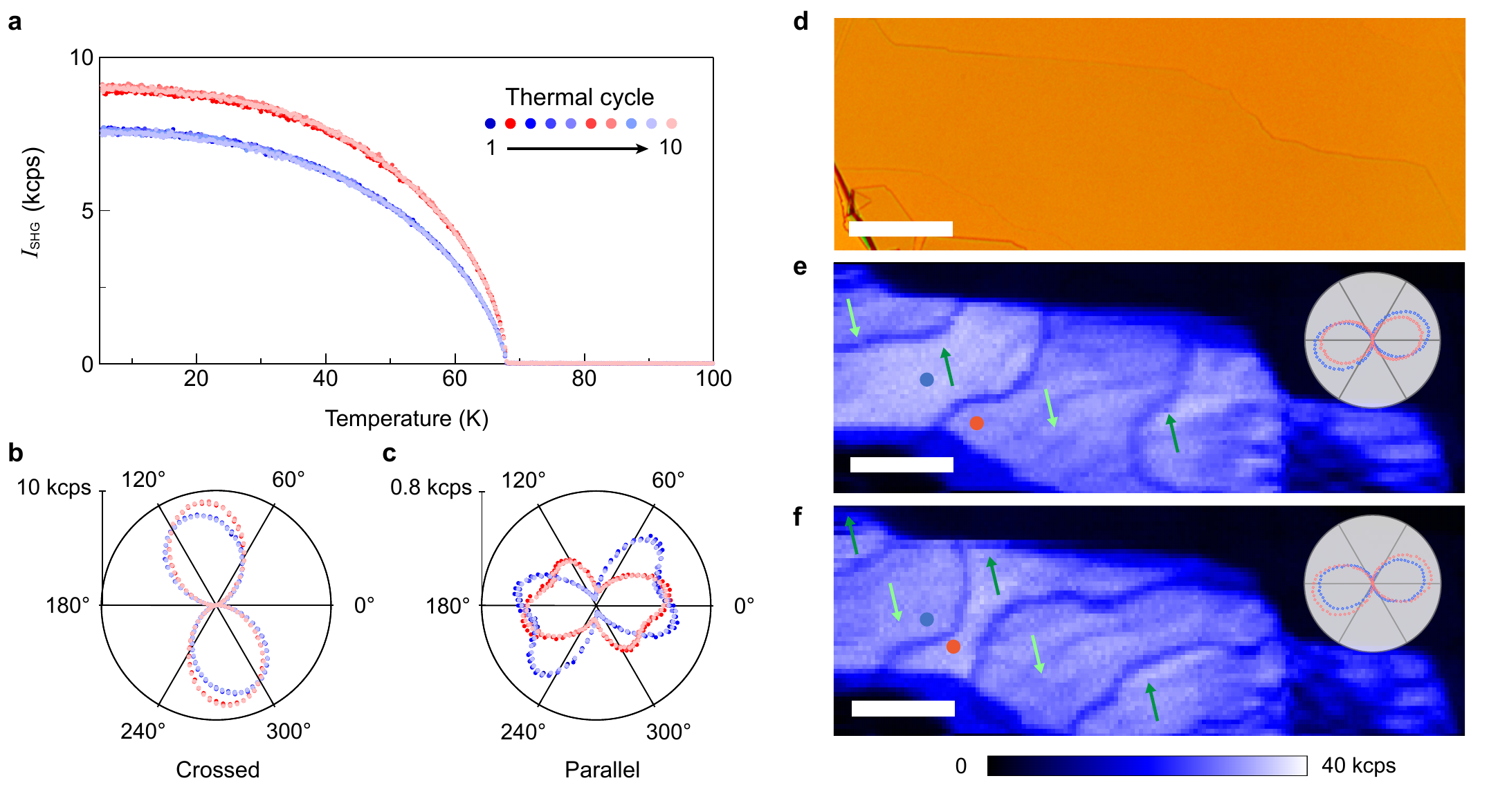}
\caption{\textbf{Ising-type N\'eel vector switching in MnPSe$_3$} {\bf a}, Temperature-dependence of SHG intensity at the same spot on a $\sim$15-$\mu$m  thick flake for 10 consecutive cooling runs across $T_N$. {\bf b-c}, Polarization-dependent SHG patterns ({\bf b}: crossed and {\bf c}: parallel) measured at 5 K after each cooling in {(\bf a)}. {\bf d}, An optical image of a thick MnPSe$_3$ flake ($\sim$100 nm). Scale bar: 50 $\mu$m. {\bf e-f}, The SHG spatial mapping at 5 K after the first cooling ({\bf e}) and the second cooling ({\bf f}) on the thick sample.  Light and dark green arrows show the direction nodes in the crossed polar pattern, which is close to the N\'eel vector direction. Scale bar: 50 $\mu$m. Inset: crossed patterns on two sides of a domain wall with positions marked by orange and blue dots.}
\label{Fig2}
\end{figure*}

After confirming the detection of AFM order by SHG, we use scanning SHG microscopy with 2 $\mu$m spatial resolution to study the AFM domains. In MnPSe$_3$ with in-plane spins and large XY anisotropy \cite{jeevanandamjpcm1999, wiedenmannssc81}, six energetically equal magnetic domains are expected due to  the threefold rotational crystalline anisotropy and the time-reversal operation \cite{oshikawaprb00, louprl07,cheongnpj19}.  Nevertheless,   Fig. \ref{Fig2}\textbf{a} shows that temperature-dependent SHG under ten consecutive cooling runs across $T_N$ on the same spot of a 15-$\mu$m thick sample collapses  onto two curves instead of six. The crossed and parallel polar patterns that respond to these two AFM domains are shown in Fig. \ref{Fig2}\textbf{b,c}.  To figure out the relation between these two domains, we performed spatial scanning  SHG microscopy at 5 K with angles of the two polarizers chosen near the maximum signal in the crossed pattern. An optical image of a region with uniform thickness ($\sim$ 100 nm, a second exfoliated thick sample) and SHG maps at 5 K after two cooling processes across  $T_N$ are shown in Fig. \ref{Fig2}\textbf{d-f}. Sharp dark lines with very low SHG intensity are observed, with bright domains of high and nearly equal SHG intensity on both sides.   In one of the regions, we pick up a few points such as the blue dot in Fig. \ref{Fig2}\textbf{e} and observe the same polar patterns shown as blue in the top right of Fig. \ref{Fig2}\textbf{e}. Crossing the dark line to a different region, we pick up a few points such as the orange one and observe the crossed polar patterns rotated by a small angle shown as orange on the top right of Fig. \ref{Fig2}\textbf{e}. By keeping the laser spot  at the orange point and performing a few thermal cycles, the polar pattern switches only between the blue and orange ones shown in  Fig. \ref{Fig2}\textbf{e}.  We interpret the two regions with high SHG intensity as two different AFM domains where the spins are reversed by 180$^\circ$ under the time-reversal operation and the dark lines are domain walls due to destructive interference \cite{fiebigreview05, yinsci2014}. The arrows in Fig. \ref{Fig2}\textbf{e,f} indicate the opposite directions of the N\'eel vectors in different regions determined by SHG polar pattern measurements at 5 K. (See Supplementary Note 1.) A second SHG map after a thermal cycling across $T_N$ in Fig. \ref{Fig2}\textbf{f} shows that the domain wall is not pinned and different regions still only have the two kinds of polar patterns shown in \ref{Fig2}\textbf{e}. (See the mapping on a 30 $\mu$m-thick naturally grown sample in Supplementary Figure 6.) The reason why we could observe the domain wall between two AFM regions with a $\pi$ phase shift by the destructive SHG interference is that the SHG has both ED and EQ contributions and only the ED term is sensitive to the $\pi$ phase shift. One could write the signal we observe as $I_i(2\omega) \sim  |\sum_{jk} \pm \chi^{ED}_{ijk}(\mathbf{L})  E_j(\omega) E_k(\omega) + \sum_{jkl} \chi^{EQ}_{ijkl}E_j(\omega)\nabla_kE_l(\omega) |^2$, where the $\pm$ signs indicate the sign change of the ED term under time-reversal operation.  As shown in Fig. \ref{Fig1}\textbf{b} and \ref{Fig2}\textbf{a}, the ED contribution at 5 K is $\sim$ 200 times larger than the EQ part and therefore the two AFM domains have high and nearly equal SHG while the domain wall has very low SHG with the EQ contribution only.

\begin{figure*}
\centering
\includegraphics[width=\textwidth]{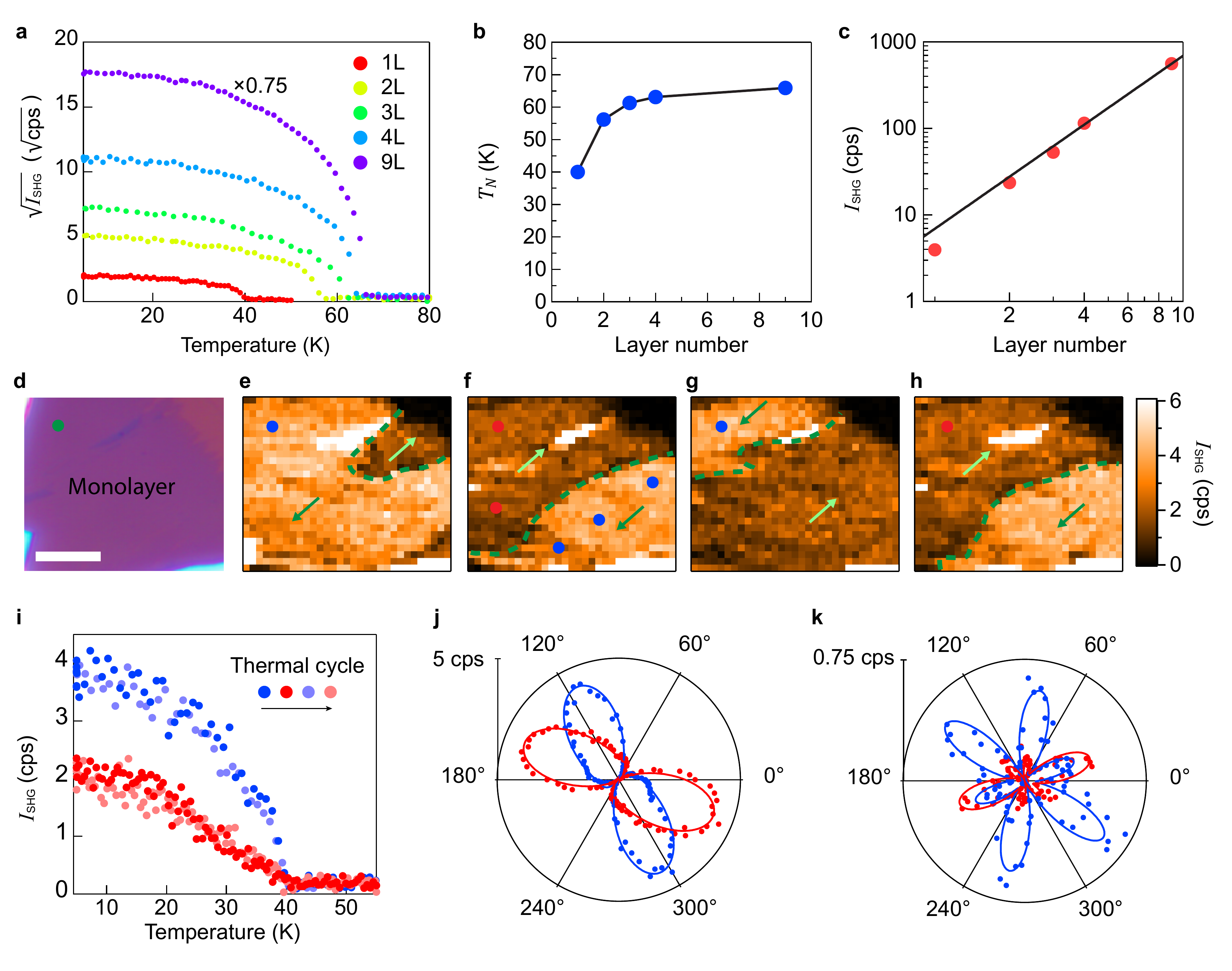}
\caption{\textbf{SHG and N\'eel vector switching of atomically-thin MnPSe$_3$ samples exfoliated in a glove box.} {\bf a}, Temperature-dependent SHG response on samples with different layers. {\bf b}, Layer-dependent N\'eel temperature. {\bf c}, Layer-dependent SHG intensity in a log-log plot. The solid line is a fit for $I\sim N^2$. {\bf d}, Optical image of a monolayer sample (S1). Scale bar: 10 $\mu$m. {\bf e-h}, SHG intensity mapping over the monolayer sample at 5 K after different thermal cycles. The dark and the light regions represent two different domains where the N\'eel vector switches by 180$^\circ$. The domain walls are highlighted by the green dashed lines. The N\'eel vectors in two domains are denoted by the green arrows. {\bf i}, Temperature-dependent SHG intensity measured at the point marked by green in ({\bf d}). Curves from four different thermal cycles are shown. {\bf j-k}, SHG polar patterns ({\bf j}: crossed and {\bf k}: parallel) for the two domains marked by red and blue, respectively.  SHG polar patterns at the selected red and the blue dots in ({\bf e-h}) are confirmed to have the same crossed pattern with the same color in ({\bf j}). See raw data in Extended Data Figure 4.}
\label{Fig3}
\end{figure*}

\begin{figure*}
\centering
\includegraphics[width=\textwidth]{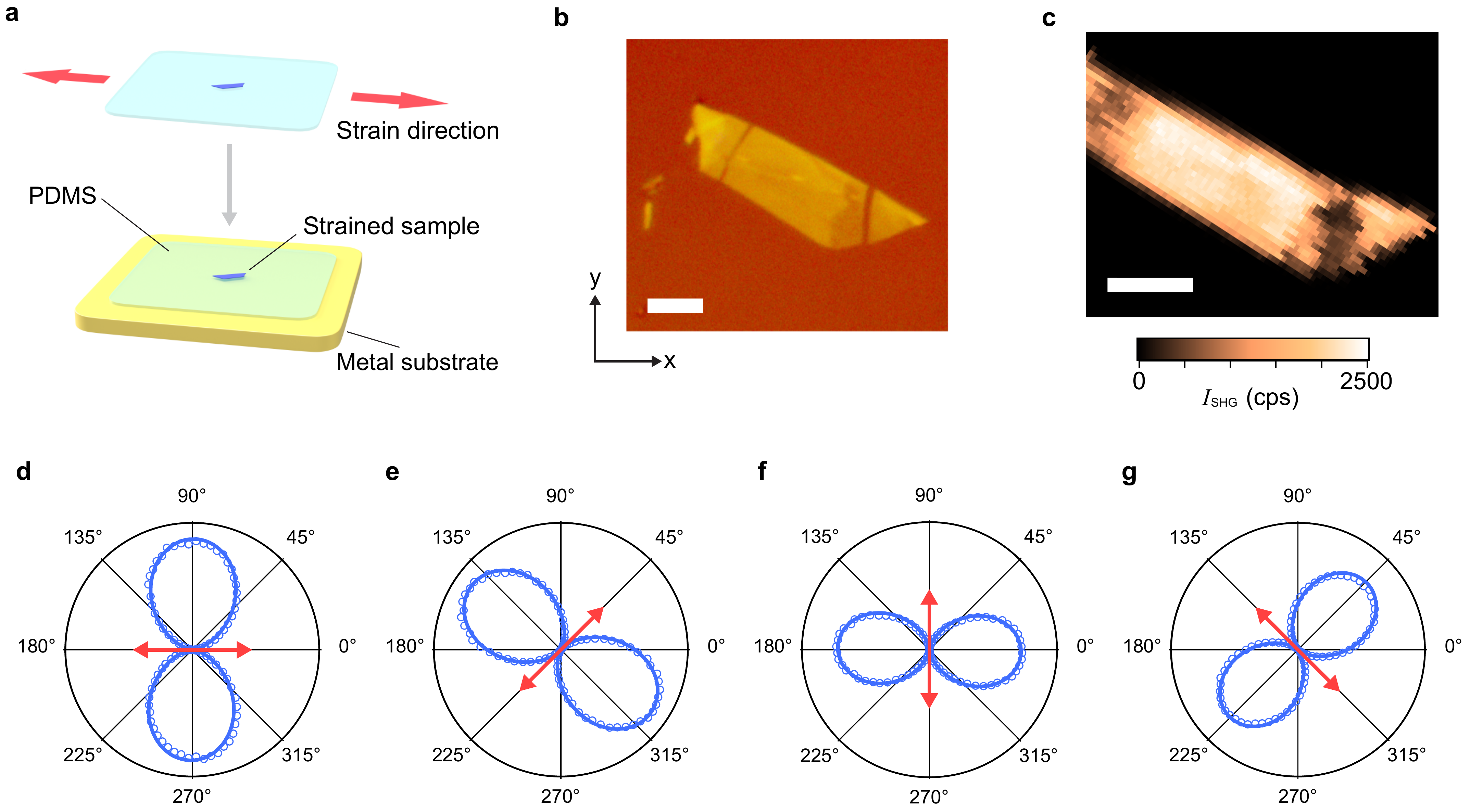}
\caption{\textbf{Strain-tunable N\'eel vector in MnPSe$_3$.} {\bf a}, A schematic of the method to control the in-plane strain. A MnPSe$_3$ flake is first exfoliated on a PDMS substrate. A tunable stretching force is then applied by a micro-manipulator on the PDMS to transfer strain to the sample, which is maintained by attaching the PDMS to a gold-coated sample holder. {\bf b}, An optical image of a stretched 15 nm-thick MnPSe$_3$ flake with a uniaxial $\sim$ 2 \% along the $x$ axis. Scale bar: 10 $\mu$m. {\bf c},  SHG spatial mapping with a 2 \% strain applied along $x$ axis at  nominally 5 K.  Scale bar: 10 $\mu$m. {\bf d}-{\bf g}, Crossed polar patterns with strain direction along ({\bf d}) 0$^{\circ}$, ({\bf e}) 45$^{\circ}$, ({\bf f}) 90$^{\circ}$  and ({\bf g}) 135$^{\circ}$  with respect to the $x$ axis. The red arrow represents the strain direction. The measured data are shown in blue open circles with fits shown in blue lines.}
\label{Fig4}
\end{figure*}

Before we discuss the origin of the two-state Ising order instead of the six-state clock order, we investigate whether the AFM order exists and direct imaging of N\'eel vector switching could be detected in the monolayer first. The ultra-thin flakes down to the monolayer are exfoliated on  SiO$_2$/Si wafers. The number of layers are determined by a combination of atomic force microscopy and optical contrast measurements \cite{casiraghinanoletter07}. (See Extended Data Figure 1.)  To probe the intrinsic properties of the materials, we exfoliate samples down to the monolayer in a glove box. Fig. \ref{Fig3}\textbf{a} shows the layer-dependent square root of the SHG intensity measured as a function of the temperature.   All of the thin flakes show a clear phase transition down to the monolayer with the layer-dependent transition temperature shown in Fig. \ref{Fig3}\textbf{b}. The transition temperature decreases from 66 K in the nine-layer sample to 56 K in the bilayer sample and it is 40 K measured in three different monolayer samples (see Extended Data Figure 2-3 for the other two monolayer samples). The intensity of SHG at 5 K from 1 layer to 9 layers nearly follows a quadratic dependence on the layer count (Fig. \ref{Fig3}\textbf{c}), which is expected for breaking the inversion symmetry in all these samples \cite{zhaolight16, liusci20}. This is different from synthetic layered AFM CrI$_3$ which supports SHG signals only with even numbers of layers \cite{sunnat19}.     Fig. \ref{Fig3}\textbf{d} shows an optical image of the monolayer   MnPSe$_3$ sample 1 (S1).   We performed twelve thermal cycles at the green dot shown in Fig. \ref{Fig3}\textbf{d} with the temperature-dependent SHG collapsed on two curves, and we plot four of them as examples in Fig. \ref{Fig3}\textbf{i}. The corresponding polar patterns for the two AFM states are shown in Fig. \ref{Fig3}\textbf{j,k}. Note that the change of the orientation in the crossed patterns in the monolayer is larger than that in bulk samples because of the reduced intensity ratio between the ED and EQ terms with reduced thickness. Fig. \ref{Fig3}\textbf{e-h} shows four SHG maps at 5 K after thermal cycles and the green dashed lines are mobile AFM domain walls.  All of them exhibit a contrast with two domains, with Fig. \ref{Fig3}\textbf{e} showing a dominant bright domain. Measured polar patters on selected  dots marked by red and blue in  Fig. \ref{Fig3}\textbf{e-h} have the same red and blue patterns shown in Fig. \ref{Fig3}\textbf{j,k}. (See Extended Data Figure 4 for more data.) We would like to point out that we observed a drop of $T_N$ by 2-4 K in the bilayer and by 18 K in the monolayer  due to aging effects when samples are exfoliated in air. (See Extended Data Figure 5.)

We observed the same two-state Ising order on all of the samples with thickness from monolayer to $\sim$ 30 $\mu$m either exfoliated on SiO$_2$/Si or directly glued on the metal platform of the cryostat. All of them show two-fold crossed polar patterns, regardless whether they are prepared in air or in a glovebox. (See Extended Data Figure 2-6, Supplementary Figure 2, 4, 7 and 9.) Since the flakes are exfoliated on SiO$_2$ and the bulk $\mu$m-thick crystals  are glued on the metal platform directly, a certain strain amount is inevitable. Therefore, we hypothesized that the Ising anisotropy is induced by the strain. In order to verify it, we deliberately apply a $\sim$ 2 $\%$ uniaxial strain by exfoliating a 15 nm flake on the polymer polydimethylsiloxane (PDMS) \cite{liunatcomm14, zhangafm16} and then stretch the PDMS as shown in Fig. \ref{Fig4}\textbf{a}. The strain strength is determined by measuring the length of the optical image of the sample along the elongation direction shown in  Fig. \ref{Fig4}\textbf{b} before and after stretching (see Supplementary Figure 12 and Supplementary Note 4). The SHG mapping with $\sim$ 2 $\%$ strain along the $x$ direction, shown in Fig. \ref{Fig4}\textbf{c}, is quite homogeneous and the crossed polar patterns at different positions all point along the same direction (Fig. \ref{Fig4}\textbf{d}.), which indicates that the strain aligns the N\'eel vector. We further apply the strain along 0, 45, 90 and 135 degrees with respect to the $x$ axis defined in Fig. \ref{Fig4}\textbf{b} and found that the crossed polar pattern follows the rotation of the strain as shown in Fig. \ref{Fig4}\textbf{d-g} within the experimental accuracy of $\pm$ 10$^{\circ}$, which indicates that the N\'eel vector is locked to the strain. We also demonstrated the N\'eel vector rotation by strain in a 3L sample. (See Extended Data Figure 7.) Because PDMS is transparent and reduces color contrast, the 3L sample is almost invisible on PDMS. (See Extended Data Figure 7.) Instead of direct straining of a monolayer on PDMS, we exfoliate a monolayer sample S3 with a long wavy shape on SiO$_2$/Si substrate  to induce strain imhomogeinty in different regions, and find that N\'eel vector direction is also locked to the local strain and points to different directions in different regions. (See Extended Data Figure 3.) We also find that the parallel polar patterns are different between monolayers S1 and S2. Fittings of the patterns indicate that the relative angles between the N\'eel vector and the crystal axis are different among samples S1 and S2, indicating that strain directions in two samples are different. (See Extended Data Figure 2.)  We would like to point out that  the N\'eel vector could be rotated to any direction by the strain in atomically thin MnPSe$_3$ due to the strain-locked Ising order, which is drastically different from a non-XY system where N\'eel vectors are switched between principal crystal axes only \cite{chennatmat2019}. We also noticed that the strain does not change the transition temperature, most likely because the strain-induced anisotropy is much smaller than the large XY anisotropy in this system \cite{jeevanandamjpcm1999}. (See Extended Data Figure 8, Supplementary Figure 13 and Supplementary Note 4.)

\begin{figure}
\includegraphics[width=0.45\textwidth]{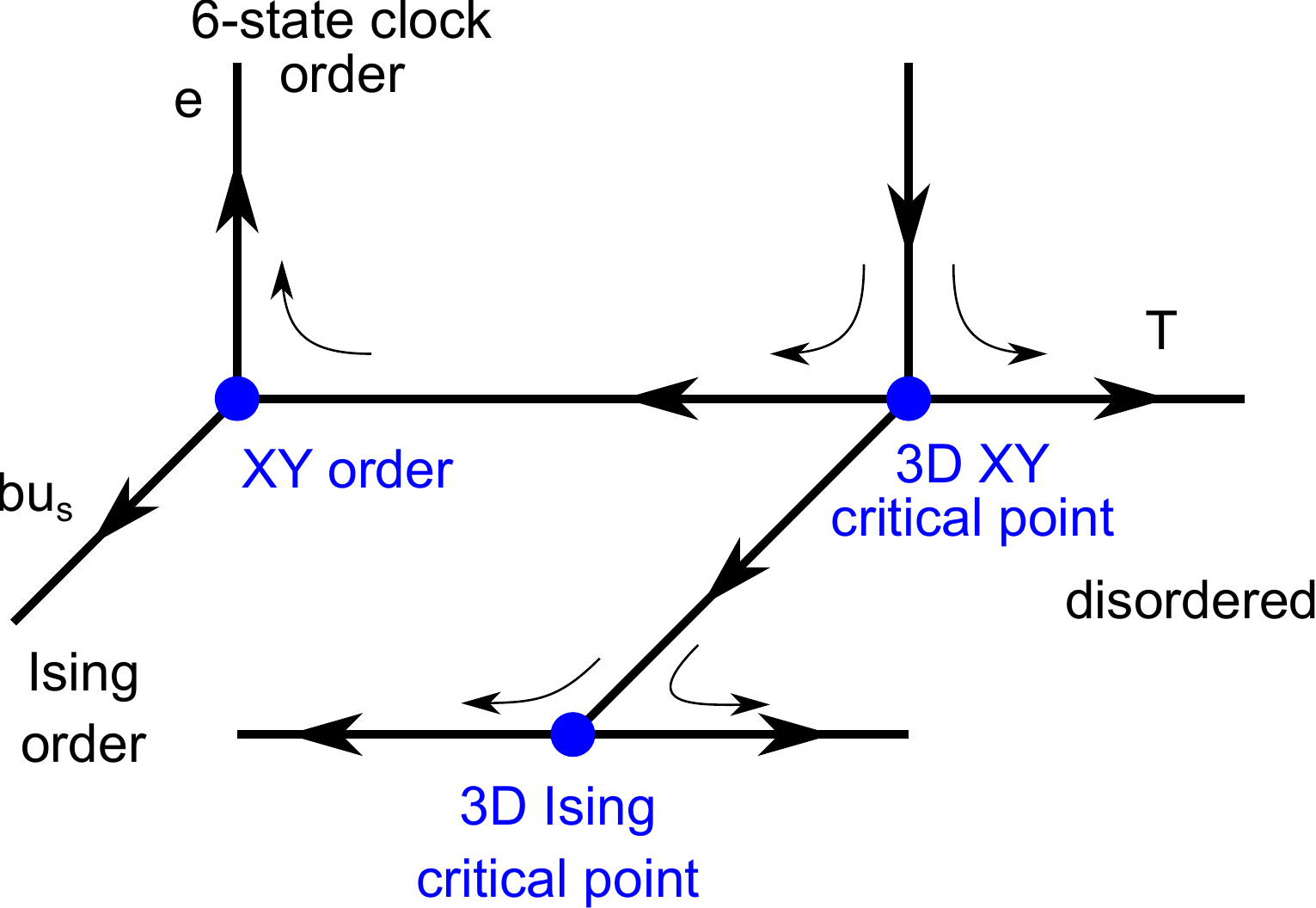}
\caption{\textbf{RG (Renormalization group) flow of the Landau theory. } When $\mu_s=0$, it flows into the 3D XY critical point and develops the six-state clock order at low temperature. When $\mu_s \neq 0$, it flows into the 3D Ising critical point and develops the two-state Ising order at low temperature. }
\label{RG}
\end{figure}

In order to understand why strain leads to an Ising order instead of a six-state clock order, we first employ a Landau expansion for free energy as a function of the N\'eel vector $L$, which applies close to the critical temperature when $L$ is small.    The lowest order term that accounts for the three-fold crystalline anisotropy, along with $\mathcal{PT}$ symmetry, is $L^6 \cos6\theta$, where $\theta$ is the polar angle measured from the $a$ axis in Fig. \ref{Fig1}\textbf{a}.    The strain is described by a second-rank tensor whose principal axes designate the directions of tensile and compressive strain.   It can be diagonalized with a rotation about the $c$ axis by an angle $\theta_0$, and it provides the term $u_s L^2 \cos 2(\theta-\theta_0)$.   Therefore, the $\theta$ dependent terms in the Landau free energy then take the form
 
\begin{equation}
F(\theta) =  b u_s L_0^2 \cos 2(\theta - \theta_0) + e L_0^6 \cos 6\theta
\end{equation}

Where $b$ and $e$ are coefficients, $u_s$ is the amplitude of the strain and $L_0$ is the magnitude of the N\'eel vector.  As shown in Fig. \ref{RG}, for $u_s=0$, in the absence of strain, this describes a six-state clock model, whose critical behavior is expected to be in the XY universality class.    For $u_s \ne 0$, the Ising anisotropy dominates, and the critical behavior is in the Ising universality class \cite{oshikawaprb00, louprl07}.   In general, the direction of the N\'eel vector will be determined by the competition between the strain and the crystalline anisotropy (See Supplementary Note 5.) This model explains why an Ising order is also induced in the $\sim$ 15 $\mu$m thick bulk sample mounted by glue as the strain is quite small.

These conclusions also follow from a more microscopic model of spins on a honeycomb lattice with interactions that reflect the symmetries of the crystal.  We consider an XY model with nearest-neighbor couplings that are modified by strain and have the form

\begin{eqnarray}
H= J_{\|,ij} {\mathbf{s}}_{i, \|}  {\mathbf{s}}_{j, \|} + J_{\perp,ij} {\mathbf{s}}_{i, \perp}   \cdot {\mathbf{s}}_{j, \perp} + J_c  \left({\mathbf{s}}_{i, \|} {\mathbf{s}}_{j, \perp} + {\mathbf{s}}_{i, \perp} {\mathbf{s}}_{j, \|} \right)
\end{eqnarray}
distinguishing the coupling for spin polarizations ``along" ($J_{\|,ij}$) and ``perpendicular to" ($J_{\perp,ij}$) the $ij$-th Mn-Mn bond and $J_c$ is a symmetry allowed cross coupling between orthogonal spin components $s_{\|,ij}$ and $s_{\perp,ij}$ in each bond.  The total energy is proportional to $|L|^2 \cos( 2 (\theta -  \theta_o - \alpha/2))$, where $\alpha/2$ defines a misalignment angle of the spin orientation from the principal strain axis. As long as the response is linear, decreasing the amount of strain while keeping the same direction will not change the tipping angle. (See Supplementary Note 6 for more details.)

The direct imaging of  N\'eel vector switching in the monolayer goes beyond previous works of domain imaging in bulk AFM samples \cite{fiebigreview05, fiebigjpd05, wadleynatnano2018} and opens the possibility of ultra-compact AFM spintronics.   Our work also creates a new method to control the 2D antiferromagntism besides electric gating \cite{huangnatnano18, jiangnatmat18, jiangnatnano18} and magnetic fields \cite {sunnat19}. Additionally, our imaging and strain-tuning methods are generically applicable to other van der Waals AFM materials including the intrinsic AFM topological insulator \cite{otrokovnature19}.  Looking forward, we hope that the discovery of extremely strain-tunable AFM order in atomically thin MnPSe$_3$   would stimulate further investigations in  designing  spatial strain patterns, to create spin pattern on demand for magnon propagation in low-dissipation terahertz spintronic devices. The monolayer MnPSe$_3$ also provides a platform to study the Kosterlitz-Thouless transition in a truly 2D XY magnet with six-state clock order if the strain could be tuned to zero by a voltage-controlled piezo-stage \cite{mutchsciadc2019}. The atomically thin MnPSe$_3$ might also be a good candidate for broadband, efficient upper-conversion as the gap is in the visible regime.

\section{Methods}

\subsection*{Sample Preparation}
Single crystals of MnPSe$_3$ were grown by the chemical vapor transport method. Elemental powders of high purity Mn, P, and Se were pressed into a pellet and sealed inside a quartz tube under vacuum. The tube was then annealed for a week at 730 $^\circ$C to form polycrystalline MnPSe$_3$ power, the composition of which was verified with powder X-ray diffraction. Crystals were then grown using the chemical vapor transport method with iodine as transport agent: 2 g of the powder and 0.4 g of iodine crystals were placed at the end of a quartz tube, which was sealed off at 13 cm length under vacuum. The sealed tube was then set in a temperature gradient of 650/525 $^\circ$C for four days to transport the starting materials placed at the hot end to the cold end. The Mn : P : Se ratio was measured to be 1.00(1) : 0.96(1) : 3.07(1) with energy-dispersive X-ray spectroscopy. The ultra-thin samples were prepared by a standard mechanical exfoliation process  on Si substrates with 90 nm thick SiO$_2$ from a few MnPSe$_3$ bulk crystals with $T_N$ $\sim$ 68 K. The samples were put into a vacuum environment after exfoliation. The total exposure time in air for samples exfoliated in a glove box is less than one minute before loading into the cryostat where the sample is under vacuum.

\subsection*{SHG microscopy}
The sample was loaded on a metal platform in a closed-cycle cryostat, and the temperature of the metal platform was controlled by a local heater, which induces only a sub-micron shift of the sample position between 5 K and 100 K. An ultrafast 800-nm Ti-sapphire laser pulse with a duration $\sim$ 50 fs at the repetition rate of 80 MHz was focused onto a 2 $\mu$m beam spot on the sample under normal incidence. A typical laser power of 200 $\mu$W was used except for the following cases. 500 $\mu$W was used for thick flakes and bulk crystals. In the bilayer sample in Extended Data Figure 6, 400 $\mu$W was used. No sample damage was observed during the measurement. The reflected SHG light was collected by an 50 $\times$ objective and reflected by a dichroic mirror or a D mirror into  a photomultiplier tube connected with a lock-in amplifier or a photon counter. The photon counter was locked to the 80 MHz in order to reduce the dark count below 0.2 counts per second (cps) without cooling the photomultiplier by cryogen. The detection sensitivity in the experiments was 0.2 counts per second. Because the parallel signal is one order smaller than the crossed signal, the polarization extinction ratio was important when measuring parallel patterns. The polarization of the fundamental light was controlled by a half-wave plate as well as a linear polarizer. The polarization of second-harmonic light was analyzed by a linear polarizer. The SHG imaging microscopy was achieved by moving the sample with three Attocube nano-positioners.

\subsection*{Strain Tuning}
We exfoliated MnPSe$_3$ on PDMS with a square shape and applied  tensile strain on two sides of the PDMS by a micro-manipulator. The stretched PDMS was then attached to a gold-coated sample platform. The strain was estimated by measuring the length change along the stretching direction in the optical image. A low transfer ratio ($\sim 13\%$) from PDMS to the sample was observed. To change the sample's strain direction, we peeled off the PDMS from the sample platform and then stretched it in another direction. To apply strain along the 45$^{\circ}$ and 135$^{\circ}$, we cut the four corners of the PDMS to form a smaller square shape in order to reduce twisting while applying strain.  The error bar of estimation of strain strength is $\pm 0.5\%$ and the error bar of the strain direction is $\pm 10^{\circ}$.

\subsection*{Symmetry analysis for SHG polar patterns}

For the SHG patterns above $T_N$, the angle dependence are described by 
\begin{equation}
I_{parallel}(2\omega,\phi)\propto \left|\chi^{EQ}_{xxzx}\cos 3\phi+\chi^{EQ}_{yxzx}\sin 3\phi\right|^2,
\end{equation}
\begin{equation}
I_{crossed}(2\omega,\phi)\propto \left|\chi^{EQ}_{xxzx}\sin 3\phi-\chi^{EQ}_{yxzx}\cos 3\phi\right|^2.
\end{equation}
$\phi$ is the angle of the incident linear polarization with respect to the $a$ axis of the crystal. Note that in the fit there is also a constant angle shift in $\phi$, which is the angle between the horizontal axis in the lab and the crystal $a$ axis. 
For the SHG patterns below $T_N$, we fit the crossed polar patten by 
\begin{equation}
I_{crossed}(2\omega,\phi,\theta)\propto L^2\left|\sin{(\phi-\theta)}\right|^2,
\end{equation}

\noindent and denote the node direction in the polar pattern as the N\'eel vector direction. $\theta$ is the angle of the direction of the N\'eel vector with respect to the crystal $a$ axis. See Supplementary Note 1 for more details.

\section{Acknowledgement}
We thank S.W. Cheong and O. Tchernyshyov for helpful discussions.  This project is mainly supported by L.W.'s startup package at the University of Pennsylvania. The development of the SHG photon counter was supported by the ARO YIP award under the Grant W911NF1910342 (L.W.). The measurement by the atomic force microscopy was supported by the ARO MURI award under the Grant W911NF2020166 (L.W.). The acquisition of the oscillator laser for the SHG experiment was supported by NSF through Penn MRSEC (DMR-1720530). E.J.M. acknowledges support from NSF EAGER 1838456. C.L.K was supported by a Simons Investigator grant from the Simons Foundation. D.G.M acknowledges support from the Gordon and Betty Moore Foundation’s EPiQS Initiative, Grant GBMF9069. H.W. and X.Q.  acknowledge support from NSF DMR-1753054 and Texas A\&M University President's Excellence Fund X-Grants Program. B.X and C.B. are supported by the Schweizerische Nationalfonds (SNF) by Grant No. 200020-172611. The DFT calculations were conducted with the advanced computing resources provided by Texas A\&M High Performance Research Computing.

\section{Author Contribution}
L.W. conceived the project and coordinated the experiments and theory.  L.W. designed the SHG imaging setup and built it  with Z.N.. Z.N. performed the experiments and analyzed the data under the supervision of L.W..  L.W., Z.N., E.M., and C.K. discussed and interpreted the data.   E.M. performed the spin model calculation. C.K. performed the Landau theory calculation.  A.H. and D.M. grew the crystals and performed magnetization measurements.  H.W. and X.Q. performed first-principle calculation. B.X. and C.B. performed the optical conductivity measurement. L.W. and Z.N. wrote the manuscript from input of all authors. All authors edited the manuscript.

\section{Addendum}

\textit{Data availability:} All data needed to evaluate the conclusions in the paper are present in the paper and the Supplementary Information. Additional data related to this paper could be requested from the authors.

\textit{Competing Interests: }The authors declare that they have no
competing financial interests.

\textit{Correspondence: }Correspondence and requests for materials
should be addressed to L.W. (liangwu@sas.upenn.edu)

\setcounter{figure}{0}
\renewcommand{\figurename}{{\bf{Extended Data Figure}}}

\begin{figure*}
\centering
\includegraphics[width=\textwidth]{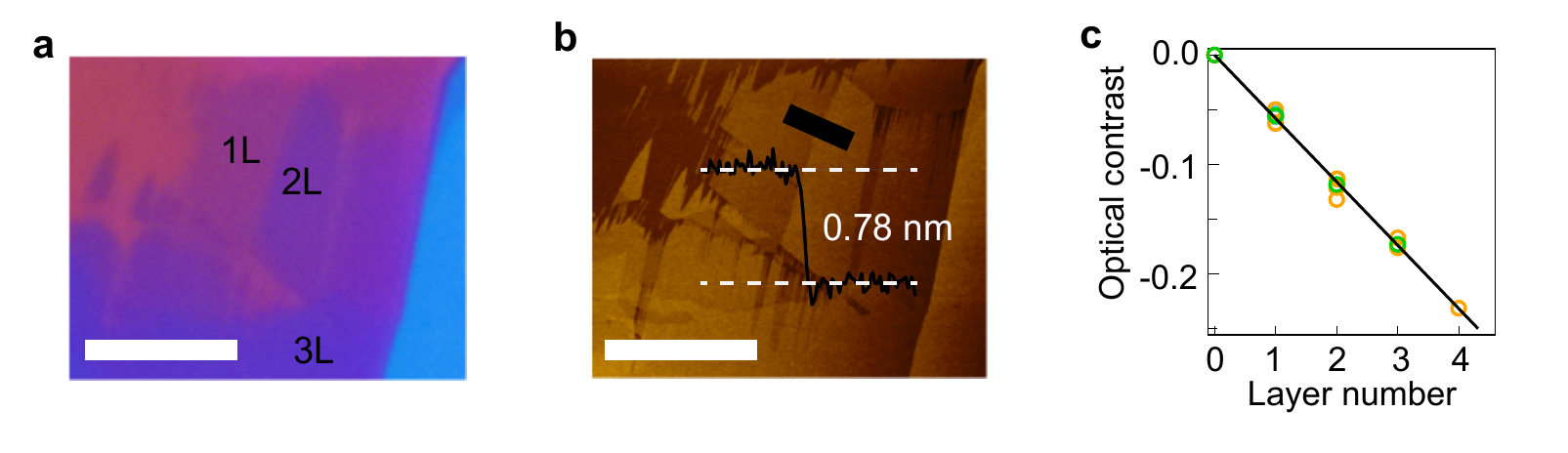}
\caption{\textbf{Thickness characterization of atomically thin MnPSe$_3$ samples.} {\bf a}, Optical image of a multilayer MnPSe$_3$ exfoliated on the 90 nm SiO$_2$/Si substrate. Scale bar: 10 $\mu$m.  {\bf b}, Atomic force microscopy image of the same sample in ({\bf a}). The step between monolayer and bilayer is measured to be around 0.78 nm. Scale bar: 10 $\mu$m. {\bf c}, Optical contrast of samples with different layer numbers. Green circles are data extracted from sample shown in ({\bf a}) and orange circles are data from other samples. The black line is a linear fit.}
\label{Extented_layerdetermination}
\end{figure*}

\begin{figure*}
\centering
\includegraphics[width=\textwidth]{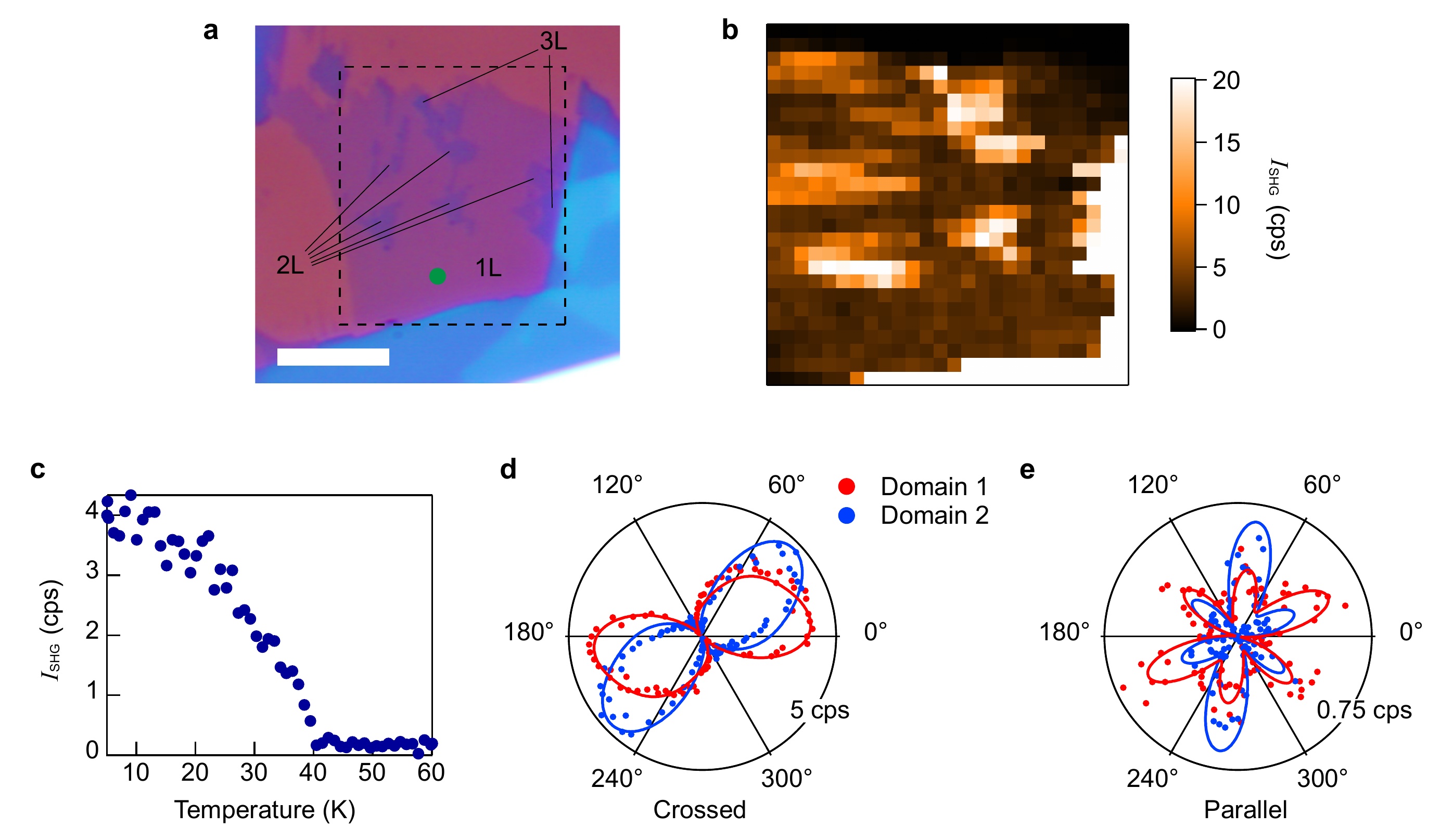}
\caption{\textbf{SHG data for monolayer S2 exfoliated in a glove box.} {\bf a}, Optical image of the monolayer S2. There are some small bilayer/trilayer islands inside the monolayer sample. Scale bar: 10 $\mu$m. The dashed region is the area for SHG mapping. {\bf b}, SHG intensity mapping of the monolayer S2. {\bf c}, SHG intensity as a function of temperature measured at the green point marked in ({\bf a}) for one thermal cycle. {\bf d-e}, Polarization-dependent SHG patterns measured at the green point in the crossed ({\bf d}) and the parallel ({\bf e}) configuration after two different thermal cycles. Data from both domains are shown. The dots are experimental data and the solid lines are the best fit. The patterns are different from the monolayer S1, indicating the angles between the N\'eel vector and the crystalline axis are different in the two samples.}
\label{ExtendedmonolayerS2}
\end{figure*}

\begin{figure*}
\centering
\includegraphics[width=\textwidth]{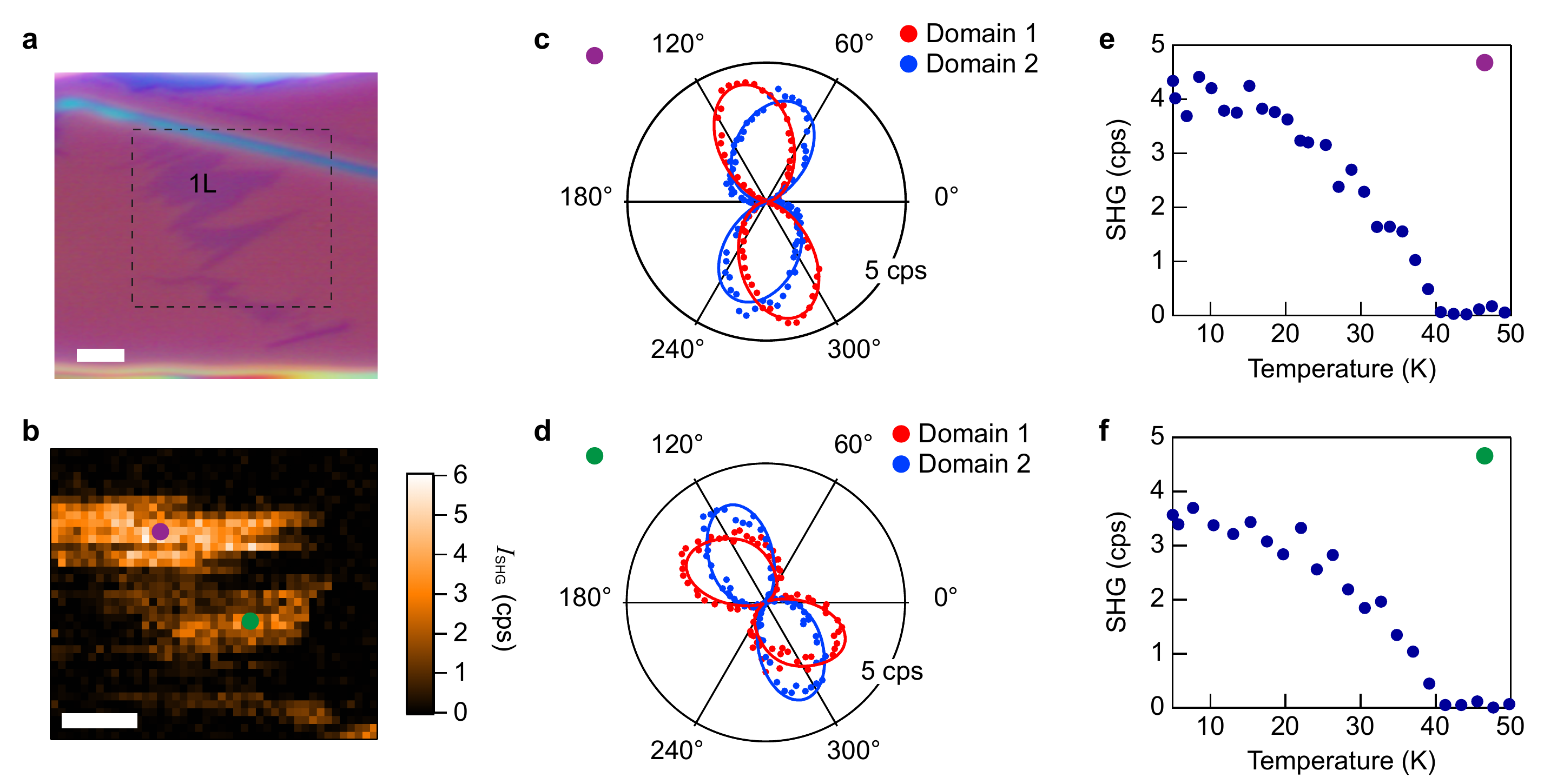}
\caption{\textbf{SHG data for monolayer  S3 exfoliated in a glove box.} {\bf a}, Optical image of the monolayer S3. Scale bar: 10 $\mu$m. The dashed box denotes the region of SHG mapping. {\bf b}, SHG intensity mapping of the monolayer S3. Scale bar: 10 $\mu$m. Polar patterns at the two points denoted by purple and green dots are measured after thermal cycles. {\bf c}, Crossed patterns of two domains of the purple point measured at 5 K after two different thermal cycles. {\bf d}, Crossed patterns of two domains of the green point measured at 5 K after two different thermal cycles. {\bf e}, SHG intensity as a function of temperature at the purple point for one thermal cycle.  {\bf f}, SHG intensity as a function of temperature at the green point for one thermal cycle. Different crossed polar patterns also with different orientations at the purple and the green points indicate that their N\'eel vectors have different directions.}
\label{ExtendedmonolayerS3}
\end{figure*}

\begin{figure*}
\centering
\includegraphics[width=\textwidth]{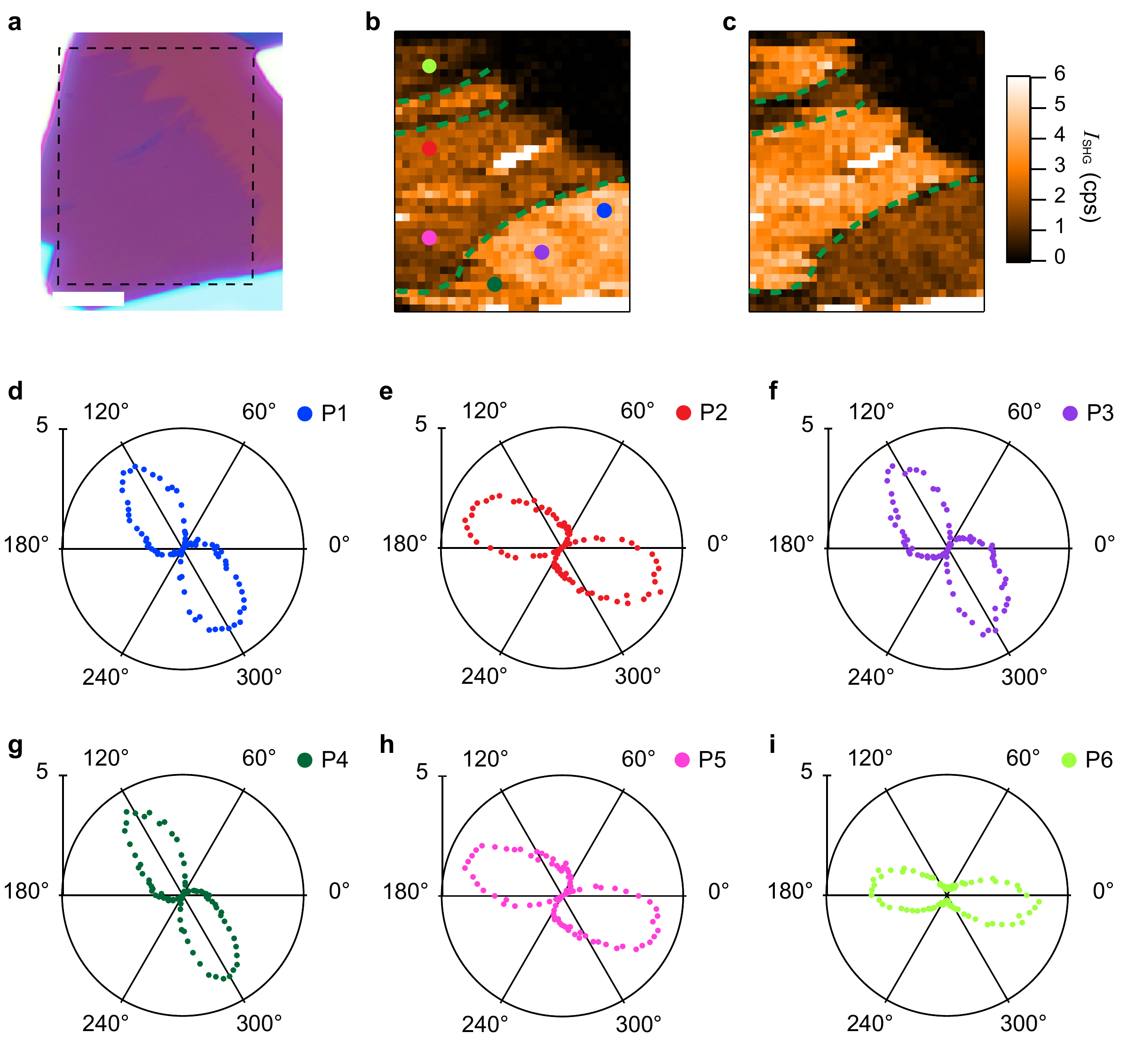}
\caption{\textbf{More data of the monolayer S1 shown in the main text.} {\bf a,} Optical image of the sample S1. Scale bar: 10 $\mu$m. The dashed region is the area for SHG mapping. {\bf b}, SHG intensity mapping at $\phi=120^\circ$ (the peak of domain 1) in the crossed pattern. {\bf c}, SHG intensity mapping at $\phi=160^\circ$ (the peak of domain 2) in the crossed pattern. Maps in ({\bf b,c}) are after the same thermal cycle. The domain walls are highlighted by the green dashed lines. {\bf d-i}, Crossed patterns of six different points marked by different colors in ({\bf{b}}). The measurement is performed at 5 K. P1-P5 are the same points in Fig. 3f in the main text. P6 is very close to sample corner and displays smaller SHG intensity with a slightly different orientation, which indicates the N\'eel vector at P6 has a slightly different direction.}
\label{Extende_monolayerS1}
\end{figure*}

\begin{figure*}
\centering
\includegraphics[width=\textwidth]{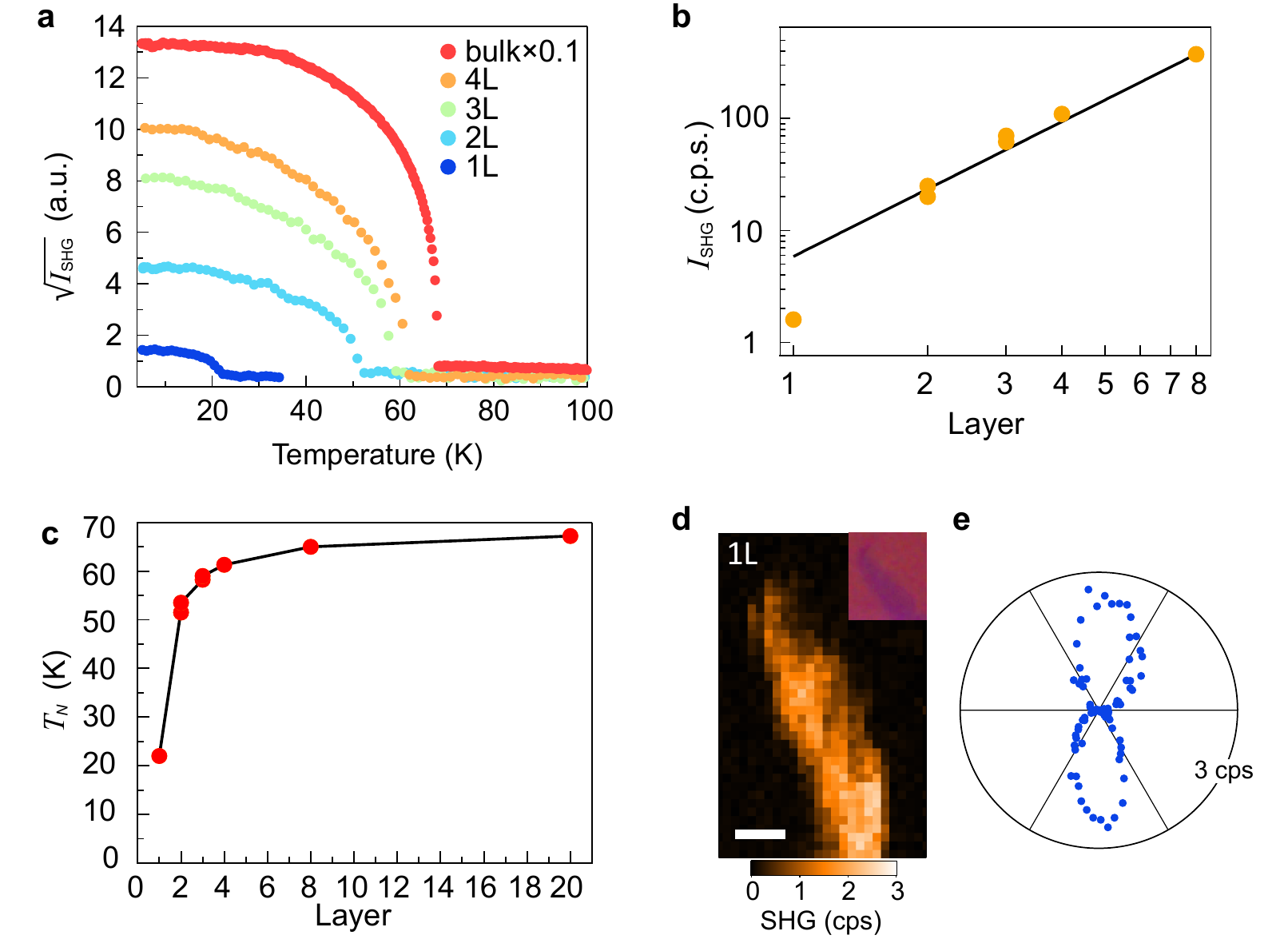}
\caption{\textbf{SHG data of atomically thin samples exfoliated in air.} {\bf a}, Square root of the SHG intensity as a function of temperature in samples with different thickness. All of the samples in this figure are exfoliated in air.  {\bf b}, SHG intensity at 5 K as a function of layer numbers in a log-log plot. Data are shown in yellow dots. The solid line is a fit for $I\propto N^2$. {\bf c}, N\'eel temperature as a function of layer numbers.  Note that there is a large reduction of both SHG signal and N\'eel temperature of monolayer compared to those exfoliated in the glove box due to the aging effect. See more data on the aging effect on a bilayer sample in Supplementary Figure 10. {\bf d}, SHG intensity mapping of a monolayer MnPSe$_3$ exfoliated in air at 5 K. Scale bar: 3 $\mu$m. Inset: Optical image of the monolayer sample. {\bf f}, Crossed polar pattern measured at the center of the monolayer sample shown in {\bf d} at 5 K.}
\label{Extended_sampleinair}
\end{figure*}

\begin{figure*}
\centering
\includegraphics[width=\textwidth]{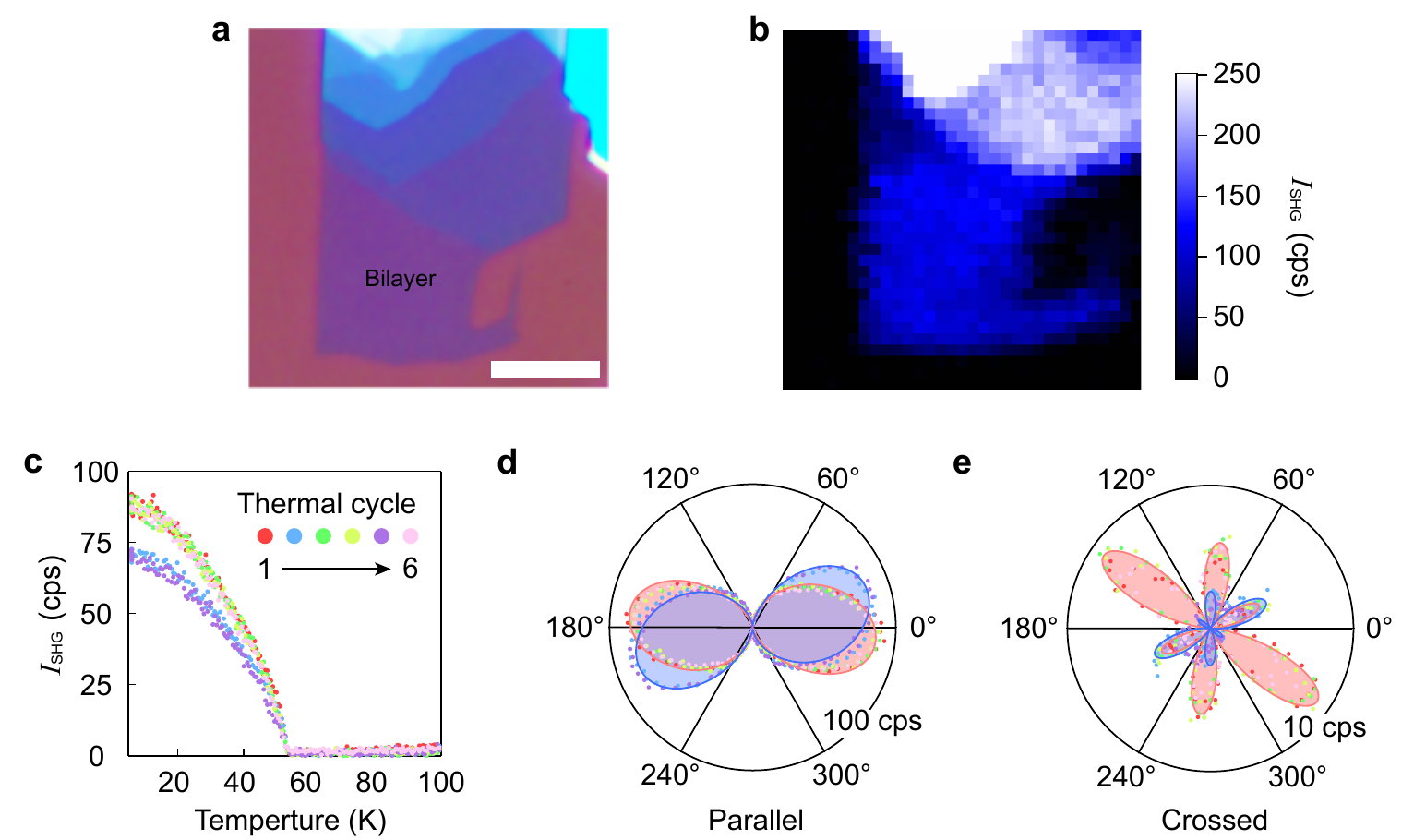}
\caption{\textbf{SHG data for a bilayer sample exfoliated in air.} This sample is exfoliated in air and different from the sample shown in the main text, which is exfoliated in a glove box. {\bf a}, Optical image of the bilayer sample. Scale bar: 10 $\mu$m. {\bf b}, SHG intensity mapping of the bilayer sample at 5 K. Note that  the SHG intensity of this sample is quite uniform, which indicates that the N\'eel vector direction is nearly aligned.  {\bf c,} SHG intensity of 6 consecutive cooling runs across $T_N$. The curves collapse into two, indicating the existence of two domains. {\bf d-e}, Crossed ({\bf d}) and parallel ({\bf e}) polar patterns measured after each cooling at 5 K. The measured data are shown in dots. The blue and red shaded regions are guides for the eye, corresponding to the two different AFM domains.  }
\label{Extend_bilayerinair}
\end{figure*}

\begin{figure*}
\centering
\includegraphics[width=\textwidth]{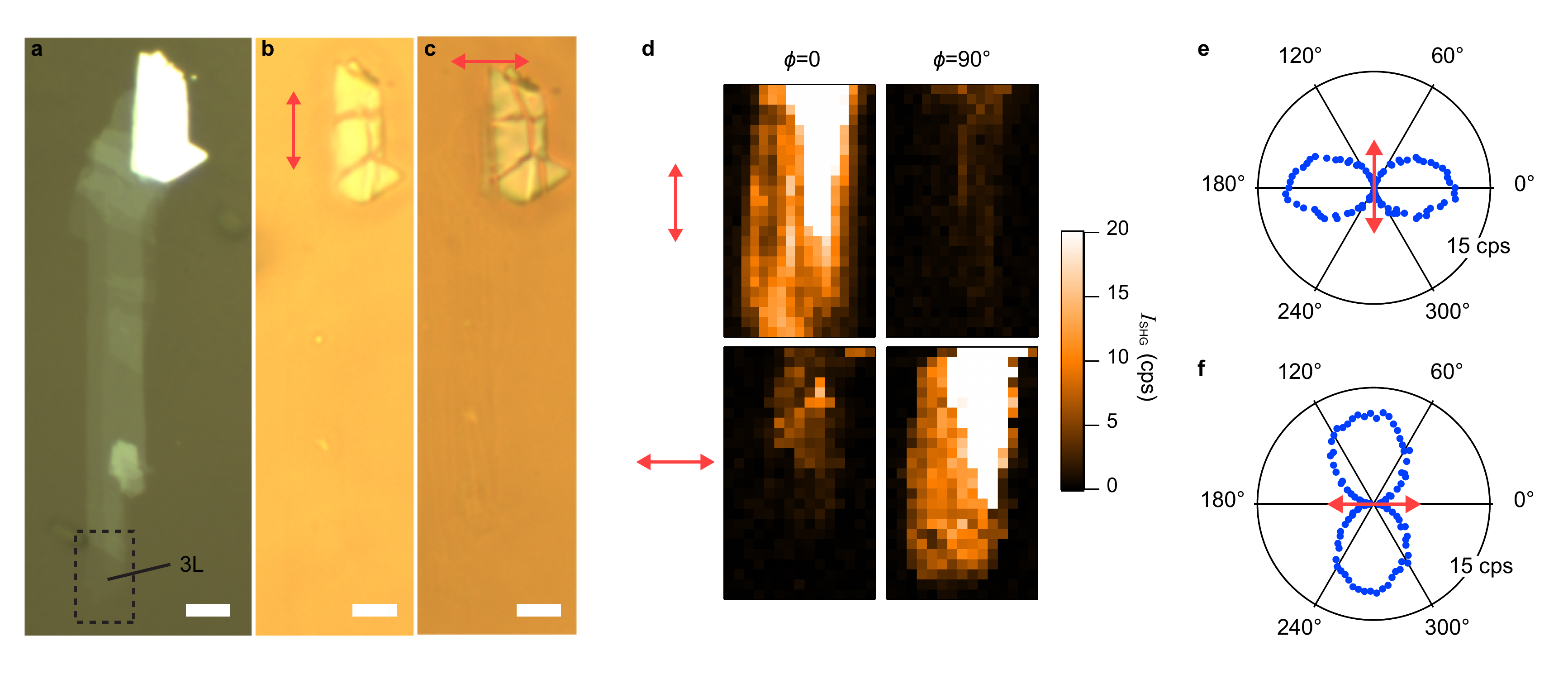}
\caption{\textbf{Strain tunability of N\'eel vector of a trilayer sample.} The strained sample is prepared using the same method as in the main text. The measured thickness is around 3.2 nm, which is a typical thickness for a trilayer sample. The measured N\'eel temperature is also consistent with a typical trilayer sample. {\bf a}, Optical image of the sample exfoliated on the PDMS before adding strains. The trilayer sample is marked by the black box. Scale bar: 10 $\mu$m. {\bf b-c}, Optical image of the sample when vertical and horizontal strains are added (marked by the red arrows), respectively. The sample and PDMS are mounted on the metal platform and the contrast of the sample is low. A 5\% strain is added on the PDMS by a micro-manipulator. The direction of the cracks of the thick flake on the top also indicates the strain direction. {\bf d}, SHG intensity mapping of the dashed area in ({\bf a}) under different strain direction (marked by the red arrows) and $\phi$ in the crossed pattern. The dark intensity maps in ({\bf d}) indicate that the N\'eel vectors are mainly along the strain direction, and rotated by around 90$^\circ$ when the strain directoin is swithced from vertical to horizontal.  {\bf e}, Crossed pattern at the center of the trilayer sample with vertical strain direction. {\bf f}, Crossed pattern at the center of the trilayer sample with horizontal strain direction. All of the above SHG measurements are operated at 5 K.}
\label{Extende_3Lstrained}
\end{figure*}

\begin{figure*}
\centering
\includegraphics[width=\textwidth]{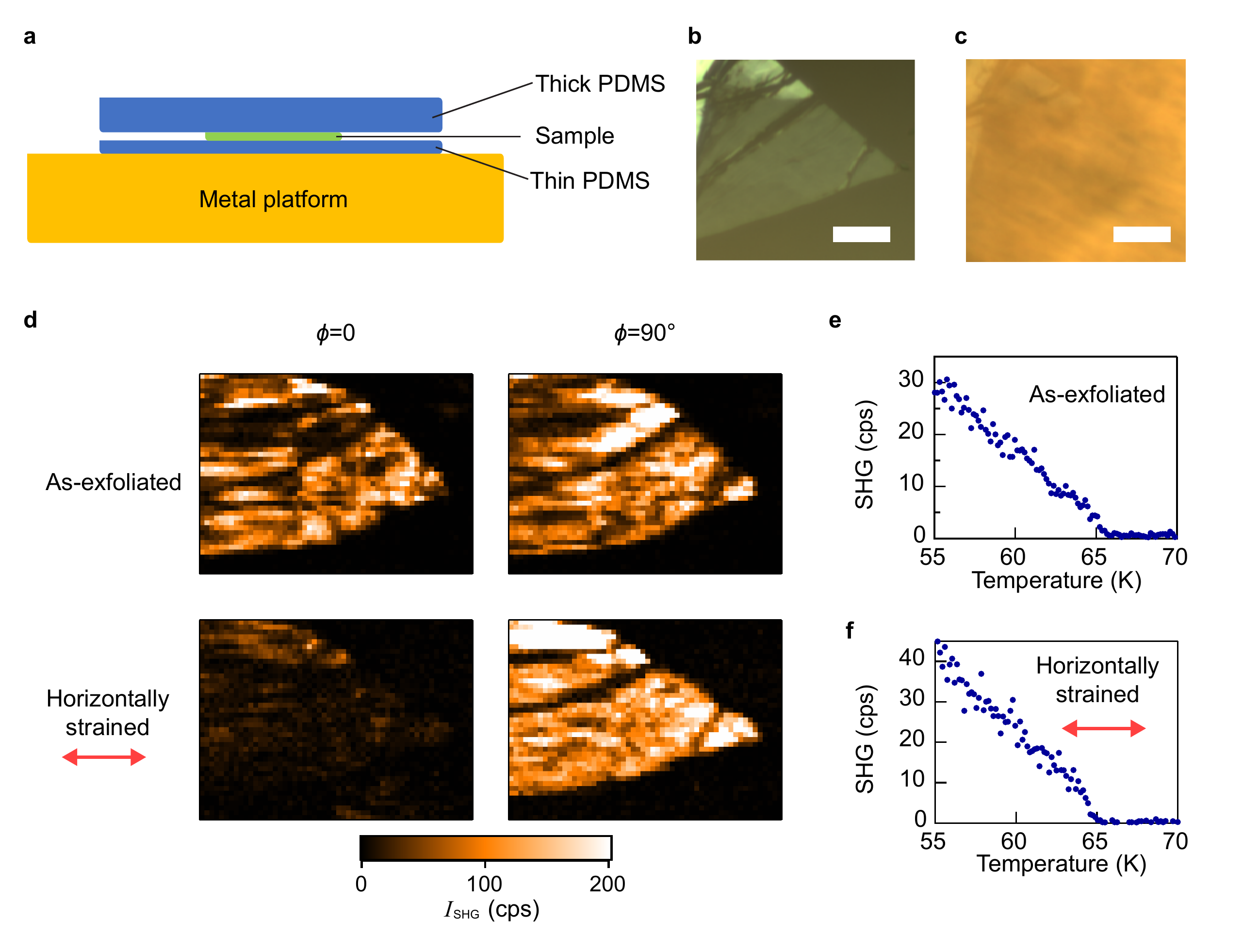}
\caption{\textbf{Strain dependence on the N\'eel temperature of a MnPSe$_3$ sample.} The MnPSe$_3$ samples in the strain-tuning experiment are exfoliated on a $\sim$30 $\mu$m PDMS, which is too thick to have the best thermal conductance. One needs to calibrate the sample temperature by measuring a thick flake on the same PDMS each time.  To measure the sample temperature without calibration, we use  thinner home-made($\sim$5 $\mu$m) PDMS to increase the thermal conductance and  encapsulate the sample after exfoliation and attach the thinner PDMS to the metal platform after applying strain (shown in {\bf a}). {\bf b,} Optical image of an as-exfoliated sample ($\sim$ 8 L) on the PDMS substrate before being put onto the metal platform. Scale bar: 10 $\mu$m. {\bf c,} Optical image of the same sample under a horizontal strain. A 5\% strain on PDMS is added by a micromanipulator and then the PDMS is attached to the metal platform. Scale bar: 10 $\mu$m. {\bf d,} SHG intensity mappings of the as-exfoliated and strained sample measured at $\phi=0^\circ$ and $\phi=90^\circ$ in the crossed polar pattern. The as-exfoliated sample shows that the N\'eel vectors are not oriented horizontally while the strained sample favors a horizontal N\'eel vector orientation. The data is collected at 5 K. {\bf e-f,} SHG intensity as a function of temperature of the as-exfoliated ({\bf e}) and strained ({\bf f}) sample after one thermal cycle. The difference of the N\'eel temperature is within 1 K.}
\label{Extended_straintemperature}
\end{figure*}

\clearpage

\begin{widetext}

\newpage

\renewcommand{\figurename}{{\bf{Supplementary Figure}}}

\setcounter{section}{0}
\setcounter{figure}{0}
\setcounter{equation}{0}
\renewcommand{\theequation}{S\arabic{equation}}

\renewcommand{\emph}[1]{{\it{#1}}}
\renewcommand{\thesection}{\arabic{section}}

\titleformat{\section}
{\normalfont\bfseries}{Supplementary Note~\thesection .}{1em}{}

\large{\textbf{Supplementary Information for ``Imaging the N\'eel vector switching in the monolayer antiferromagnet MnPSe$_3$ with strain-controlled Ising order"}}
\normalsize

\vspace{24 pt}

\section{Symmetry analysis of SHG responses}
Above the N\'eel temperature $T_N$, the bulk MnPSe$_3$ has rhombohedral (ABC) stacking and belongs to the point group $\overline{3}$ ($S_6$) and the space group 148, where all of the mirror symmetries are broken, but the threefold rotation symmetry along the $c$ axis and the inversion symmetry are present \cite{wiedenmannssc81}. Below $T_N$, the material forms N\'eel AFM order, which breaks the inversion symmetry and the threefold rotation symmetry and allows all of the 18 non-zero terms in $\chi^{(2)}_{ijk}$.

Here we consider two sources contributing to the SHG signal: an electric-quadrupole (EQ) contribution from the lattice $P^{EQ}_i(2\omega)=\sum_{jkl}\chi^{EQ}_{ijkl}E_j\nabla_kE_l$ and an electric dipole (ED) contribution related with N\'eel vector ($\bf{L}$), $P^{ED}_i(2\omega)=\sum_{jk}\chi^{(2)}_{ijk}(\mathbf{L})E_jE_k$. $\chi^{(2)}_{ijk}(\mathbf{L})$ generally depends on the N\'eel vector ($\bf{L}$). The EQ contribution is present at all temperature while the ED one is only allowed below $T_N$.

\subsection{Electric-quadrupole contribution}
First, we consider EQ contribution $P^{EQ}_i(2\omega)=\sum_{jkl}\chi^{EQ}_{ijkl}E_j\nabla_kE_l$. For a normally incident beam on the $xy$ plane, we have $E_z=0$ and $\nabla_x e^{iq_z z}=\nabla_y e^{iq_z z}=0$. The point group symmetry allows eight non-zero elements with two independent values, $\chi^{EQ}_{xxzx}=-\chi^{EQ}_{yxzy}=-\chi^{EQ}_{yyzx}=-\chi^{EQ}_{xyzy}$ and $\chi^{EQ}_{yxzx}=\chi^{EQ}_{xxzy}=\chi^{EQ}_{xyzx}=-\chi^{EQ}_{yyzy}$. Assuming the polarization of incident light is at an angle of $\phi$ with with respect to the $a$ axis, we have $E=(\cos\phi, \sin\phi, 0)$. For parallel and crossed patterns, the detected polarization is along $(\cos\phi, \sin\phi, 0)$ and $(-\sin\phi, \cos\phi, 0)$, respectively. Therefore, the SHG intensity above N\'eel temperature is
\begin{equation}
I_{parallel}(2\omega,\phi)\propto q_z^2\left|\chi^{EQ}_{xxzx}\cos 3\phi+\chi^{EQ}_{yxzx}\sin 3\phi\right|^2,
\end{equation}
\begin{equation}
I_{crossed}(2\omega,\phi)\propto q_z^2\left|\chi^{EQ}_{xxzx}\sin 3\phi-\chi^{EQ}_{yxzx}\cos 3\phi\right|^2.
\end{equation}

Both of them are sixfold rotationally symmetric with the same amplitude, and peak positions are dependent on the ratio of $\chi^{EQ}_{xxzx}$ and $\chi^{EQ}_{yxzx}$. A fit to SHG data at 100 K in a 100-nm thick MnPSe$_3$ sample is shown in Fig. 1b in the main text. Note that in the fit there is a shift of the angle by a constant set by the angle between the lab axis and the crystal $a$ axis.

\subsection{Electric-dipole contribution from the N\'eel AFM order }
Next, we consider the ED contribution $P^{ED}_i(2\omega)=\sum_{jk}\chi^{(2)}_{ijk}(\mathbf{L})E_jE_k$ related with the N\'eel vector $\mathbf{L}$. One could also write this term to relate with the N\'eel vector  by $P^{ED}_i(2\omega)=\sum_{jkl}\chi^{ED}_{ijkl}E_jE_kL_l$, where $\chi^{(2)}_{ijk}(\mathbf{L})=\sum_{l}\chi^{ED}_{ijkl}L_l$. Since all the non-zero $\chi^{(2)}_{ikl}$ are allowed in the AFM ordered state, it is impractical to fit the polar patterns to extract the components. Under normal incidence, there are still six non-zero independent terms in the ordered phase.

Consider that the crossed polar patterns from monolayer to bulk are quite similar at all temperature below the transition temperature, to derive a simpler relationship with the N\'eel vector, we use a Talyor expansion at $\mathbf{L}=0$ (near the phase transition) to expand $\chi^{(2)}_{ikl}$ up to the linear term,
\begin{equation}
\chi_{ijk}^{(2)}(\mathbf{L})=\sum_{l}\frac{\partial \chi_{ijk}}{\partial L_l}L_l|_{\mathbf{L}=0}.
\label{linear1}
\end{equation}
Then one can define a 4th-rank tensor $\chi^{N}_{ijkl}=\frac{\partial \chi_{ikl}}{\partial L_l}|_{\mathbf{L}=0}$ and the ED term is written as $P_i^{N}(2\omega)=\sum_{jkl}\chi^N_{ijkl}E_jE_kL_l$. Note that the N\'eel vector behaves the same as an electric field under the threefold rotation operation, the inversion operation, and the reflection operation vertical to the $a$ axis in the lattice of MnPSe$_3$. The former two are the symmetry operations in the lattice of a multi-layer MnPSe$_{3}$ while the last one is an additional symmetry operation in the lattice of the monolayer MnPSe$_3$ with the spins along the $a$ axis. Therefore, we could treat the N\'eel vector similar to an electric field in the symmetry analysis.
The tensor $\chi^N_{ijkl}$ has the lattice symmetry in paramagnetic phase. The non-zero components of the ED SHG susceptibility tensor in the ordered state could be derived from the symmetries of the susceptibility tensor in the high-temperature phase and the order parameter \cite{savalenti00}, $\chi^N_{ijk}=\sum_{jkl}\chi^N_{ijkl}L_l$, where $L_l$ is a component of the N\'eel vector. Note that the symmetry of $\chi^{N}_{ijkl}$ is higher than the general $\chi^{ED}_{ijkl}$ that applies to the temperature below $T_N$ because we only consider up to the linear order in Equa. \ref{linear1}. Therefore, $\chi^{N}_{ijkl}$ applies to the temperature near the phase transition and could be extended to the general ED SHG susceptibility tensor by considering higher-order terms. Note that the discussion in this paragraph applies to monolayer and bulk samples.

Monolayer MnPSe$_3$ has a mirror symmetry along $ac$ plane and a threefold rotation symmetry along $c$ axis without the AFM order. With the lattice symmetry and the exchange symmetry of indices $k$ and $l$, one can write the tensor to be
\begin{equation}
\left(
\begin{array}{cc}
\left(
\begin{array}{cc}
\chi^N_{xxxx}& 0\\
0& \frac{1}{2}\left ( \chi^N_{xxxx}-\chi^N_{xyyx} \right )\\
\end{array}
\right) & \left(
\begin{array}{cc}
0& \frac{1}{2}\left ( \chi^N_{xxxx}-\chi^N_{xyyx} \right ) \\
\chi^N_{xyyx} & 0\\
\end{array}
\right) \\
\left(
\begin{array}{cc}
0& \chi^N_{xyyx} \\
\frac{1}{2}\left ( \chi^N_{xxxx}-\chi^N_{xyyx} \right ) & 0\\
\end{array}
\right) & \left(
\begin{array}{cc}
\frac{1}{2}\left ( \chi^N_{xxxx}-\chi^N_{xyyx} \right )& 0\\
0& \chi^N_{xxxx}\\
\end{array}
\right) \\
\end{array}
\right).
\label{chi}
\end{equation}

\noindent To calculate polarization dependent SHG signal, we define $\phi$ to be the angle of incident light polarization and $\theta$ to be the angle of the N\'eel vector with respect to the $a$ axis. Therefore $E=(\cos\phi, \sin\phi, 0)$ and $L=(\cos\theta, \sin\theta, 0)$. Eventually one gets
\begin{equation}
I_{parallel}(2\omega,\phi)\propto L^2\left|\chi^{N}_{xxxx}\cos(\phi-\theta)\right|^2,
\end{equation}

\begin{equation}
I_{crossed}(2\omega,\phi)\propto L^2\left|\chi^{N}_{xyyx}\sin(\phi-\theta)\right|^2.
\end{equation}
These results enable us to identify the direction of N\'eel vector. It is interesting to note that only the orientation but not the shape and magnitude of ED-SHG patterns would change when the direction of N\'eel vector changes. 

Layer stacking in multi-layer MnPSe$_3$ breaks the mirror symmetry while maintain the threefold rotation symmetry in lattice. Therefore more tensor elements need to be considered. The new tensor is
\begin{equation}
\left(
\begin{array}{cc}
\left(
\begin{array}{cc}
\chi^N_{xxxx}& -\chi^N_{xxxy}\\
-\frac{1}{2}\left ( \chi^N_{xxxy}+\chi^N_{yxxx} \right )& \frac{1}{2}\left ( \chi^N_{xxxx}-\chi^N_{xyyx} \right )\\
\end{array}
\right) & \left(
\begin{array}{cc}
-\frac{1}{2}\left ( \chi^N_{xxxy}+\chi^N_{yxxx} \right )& \frac{1}{2}\left ( \chi^N_{xxxx}-\chi^N_{xyyx} \right ) \\
\chi^N_{xyyx} & -\chi^N_{yxxx}\\
\end{array}
\right) \\
\left(
\begin{array}{cc}
\chi^N_{yxxx}& \chi^N_{xyyx} \\
\frac{1}{2}\left ( \chi^N_{xxxx}-\chi^N_{xyyx} \right ) & \frac{1}{2}\left ( \chi^N_{xxxy}+\chi^N_{yxxx} \right )\\
\end{array}
\right) & \left(
\begin{array}{cc}
\frac{1}{2}\left ( \chi^N_{xxxx}-\chi^N_{xyyx} \right )& \frac{1}{2}\left ( \chi^N_{xxxy}+\chi^N_{yxxx} \right )\\
-\chi^N_{xxxy}& \chi^N_{xxxx}\\
\end{array}
\right) \\
\end{array}
\right).
\label{chi2}
\end{equation}

\noindent Considering a incident electric field $E_{in}=(\cos{\phi},\sin{\phi},0)$ and N\'eel vector $L=(\cos{\theta},\sin{\theta},0)$, the output second-order response is then
\begin{equation}
I_{parallel}(2\omega,\phi,\theta)\propto L^2\left|\chi^N_{xxxx} \cos{(\theta-\phi)}+\chi^N_{xxxy} \sin(\theta-\phi)\right|^2,
\end{equation}
\begin{equation}
I_{crossed}(2\omega,\phi,\theta)\propto L^2\left|\chi^N_{yxxx} \cos{(\theta-\phi)}+\chi^N_{xyyx} \sin(\theta-\phi)\right|^2.
\end{equation}

\noindent When the crossed pattern has a node (that crosses zero), according to our simulation, there are two possibilities: (1 )$\chi^N_{yxxx}$ is much smaller than $\chi^N_{xyyx}$; (2) $\chi^N_{yxxx}$ and $\chi^N_{xyyx}$ are almost in phase. In the first case, the node direction in the across pattern is the N\'eel vector direction. In the second case, their direction are off by an angle of $\arctan \frac {|\chi^N_{yxxx}|}{|\chi^N_{xyyx}|}$. To summarize, the crossed polar pattern could be described by $L^2 \sin(\theta-\phi+C)^2$, where $C$ is a constant.

\subsection{Coupling between ED and EQ terms}

In our experiment, the in-plane uniaxial anisotropy caused by the strain pins the N\'eel vector to an Ising type. Therefore, the in-plane N\'eel vector can only switch between two opposite directions reversed by the time-reversal operation. If the SHG signal is purely from the ED contribution, there is no way to distinguish these two domains because of the lack of phase information in the SHG measurement. However, by interfering the ED contribution with the EQ contribution, it is possible to get different responses from two domains even though the EQ term is much smaller than the ED term. The total SHG intensity is
\begin{equation}
I^{total}_{parallel}=\left|L\chi^N_{xxxx} \cos{(\theta-\phi)}+L\chi^N_{xxxy} \sin(\theta-\phi)+q_z\chi^{EQ}_{xxzx}\sin 3\phi+q_z\chi^{EQ}_{yxzx}\sin 3\phi\right|^2,
\label{couplingparallel}
\end{equation}
\begin{equation}
I^{total}_{crossed}=\left|-L\chi^N_{yxxx} \cos{(\theta-\phi)}-L\chi^N_{xyyx} \sin(\theta-\phi)+q_z\chi^{EQ}_{xxzx}\sin 3\phi-q_z\chi^{EQ}_{yxzx}\cos 3\phi\right|^2.
\label{couplingcrossed}
\end{equation}

In the main text, we show the exsistence of two different crossed patterns corresponding to two Ising domains. To test whether the data are consistent with the above analysis, we fit two crossed patterns of two Ising domains from a $\sim$15 $\mu$m sample using Equa. (\ref{couplingcrossed}). Crossed polar patterns for domain 1 and 2 are plotted by blue and red dots in Supplementary Figure \ref{fittingbulk}. Since the EQ term is not expected to change with temperature, we use the EQ term in the crossed pattern above N\'eel temperature. The crossed pattern measured at 100 K is shown by yellow dots in Supplementary Figure \ref{fittingbulk}d. Note the 100 K pattern is magnified by 200 times. We fit the three curves simultaneously with the fitting weight of the 100 K data being 200 times larger than the other two at 5 K. The best fit results are shown in solid lines in Supplementary Figure \ref{fittingbulk}. It matches well with all of the three patterns. The green and yellow solid lines represent the extracted pure ED and EQ contributions. In thick samples, the magnitude of the crossed pattern signal remains almost the same but with a small angle change in the node direction when domain change happens because of the much larger ED contribution than the EQ one. In thin samples, especially in the monolayer (see main text Fig. 3j), crossed patterns of two domains look more different in terms of the orientation, which results from a smaller ratio between ED term and EQ term.

\begin{figure}
\centering
\includegraphics[width=\textwidth]{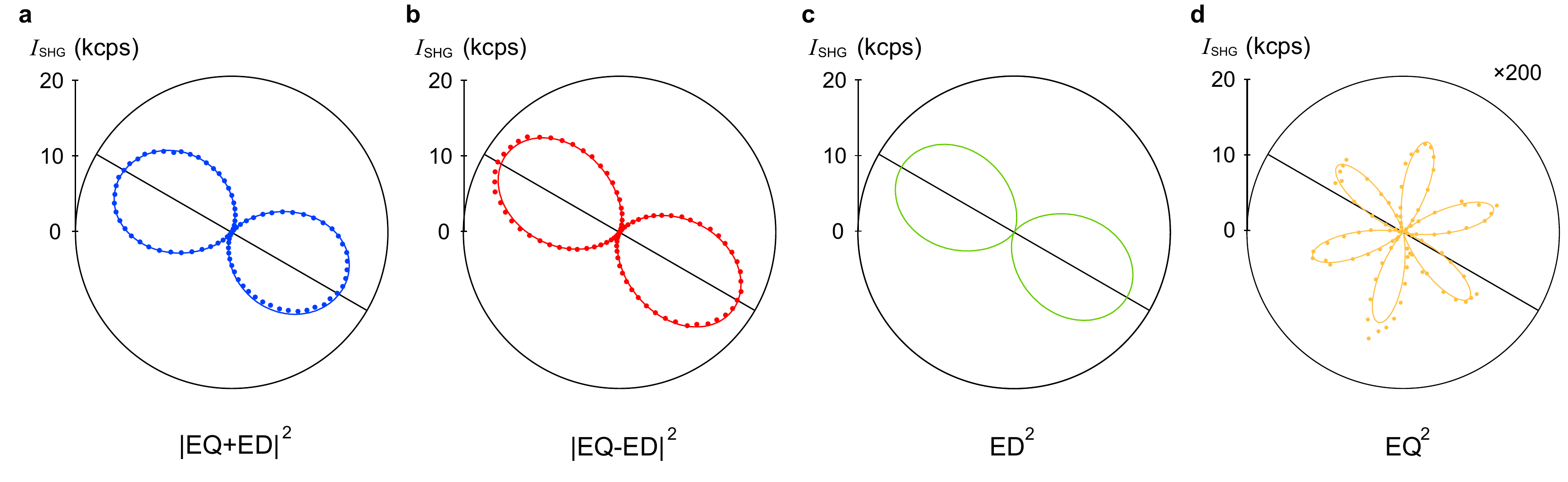}
\caption{ \textbf{A fit to crossed patterns in two Ising domains in the  $\sim$ 15 $\mu$m thick MnPSe$_3$ bulk crystal in Fig. 2a-c in the main text.} \textbf{a}, A crossed pattern of domain 1 at 5 K, \textbf{b}, a crossed pattern of domain 2 at 5 K. \textbf{c}, Extracted pure ED contribution from the fit. \textbf{d}, Crossed pattern multiplied by a factor of 200 at 100 K.}
\label{fittingbulk}
\end{figure}

The ED term in the parallel configuration at 5 K is around 4--6 times larger than the EQ term above $T_N$ (for example, see Supplementary Figure \ref{tempdependence} on a $\sim$ 50 nm sample.). It induces a significant change of the polar pattern when the N\'eel vector reverses. We have showed the switching of parallel polar patterns of a $\sim$ 15 $\mu$m thick, monolayer and bilayer samples. Here we show the domain switching in a $\sim$ 100 nm thick sample and the polar patterns corresponding to two domains are shown in Supplementary Figure \ref{parallel}. Note that this is a 3rd 100 nm sample which is different from those in Fig. 1b and Fig. 2d-f in the main text.

\begin{figure}
\centering
\includegraphics[width=0.6\textwidth]{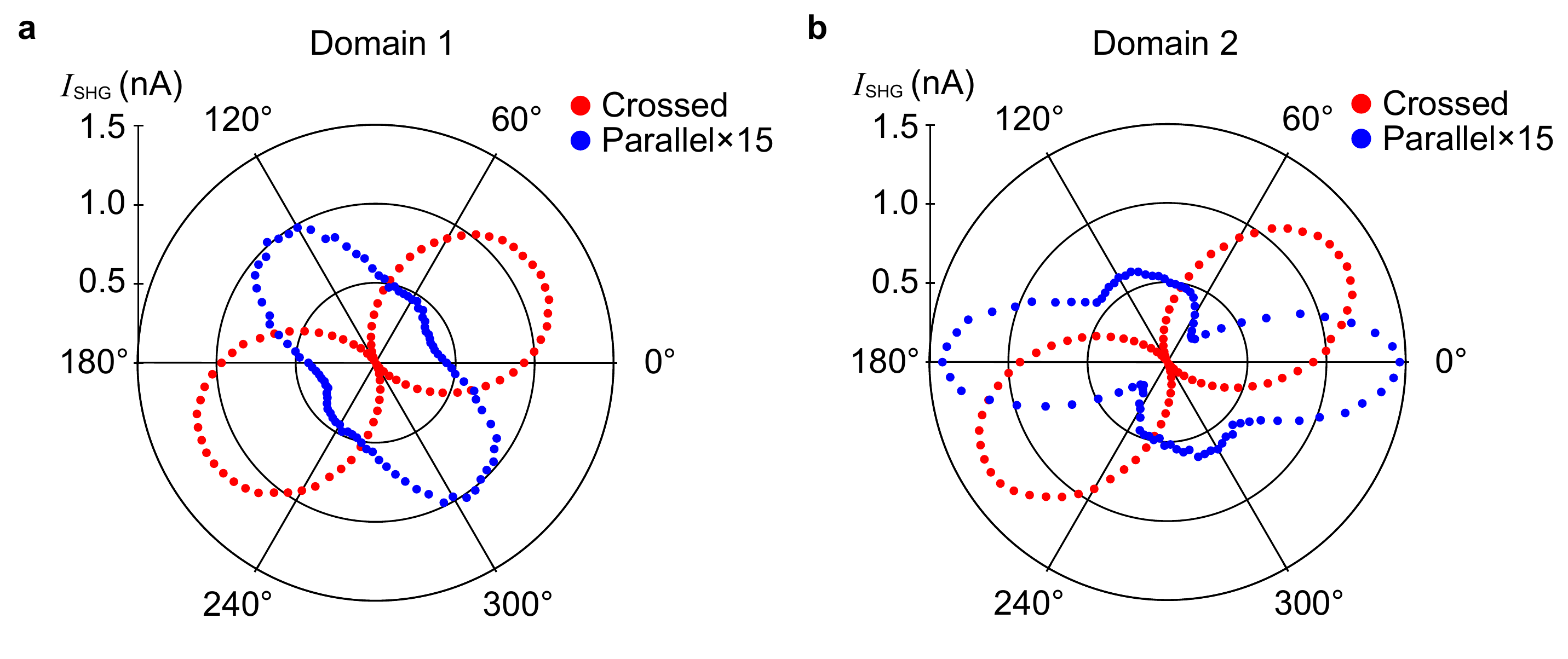}
\caption{\textbf{Polar patterns in domain switching of a third $\sim$100 nm flake on SiO$_2$/Si.} \textbf{a}, Parallel and \textbf{b}, crossed polar patterns on two domains at 5 K.}
\label{parallel}
\end{figure}

\subsection{SHG patterns in the existence of out-of-plane N\'eel vector component}
As discussed in the main text, the neutron scattering experiment reveals that spins in MnPSe$_3$ are in-plane \cite{wiedenmannssc81}. From SHG patterns, one can also tell whether the spin direction is out-of-plane or in-plane. From Equa. \ref{linear1} and Equa. \ref{chi2}, one can obtain a sixfold SHG patterns when ${\bf L}$ has a pure z-component and twofold SHG patterns when ${\bf L}$ has no z-component. When the ${\bf L}$ has both in-plane and out-of-plane components, the SHG pattern is neither twofold nor sixfold patterns. Examples of crossed patterns in each case are shown in Supplementary Figure \ref{inplaneSHG}. Based on this analysis, the twofold crossed patterns we observe in the experiment also support the picture of in-plane N\'eel vector.

\begin{figure}
\centering
\includegraphics[width=0.7\textwidth]{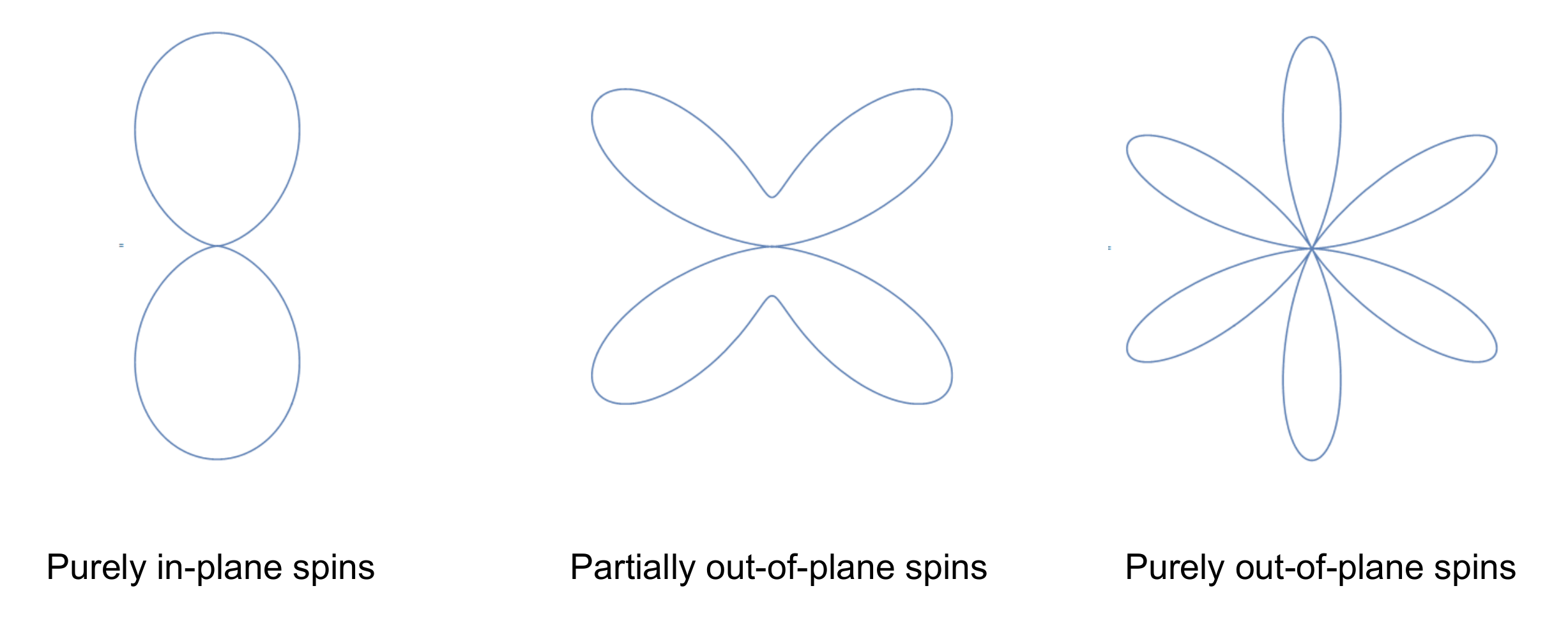}

\caption{\textbf{A comparison of crossed SHG patterns under purely in-plane, partially out-of-plane and purely out-of-plane spins.} }
\label{inplaneSHG}
\end{figure}

\begin{figure}
\centering
\includegraphics[width=0.9\textwidth]{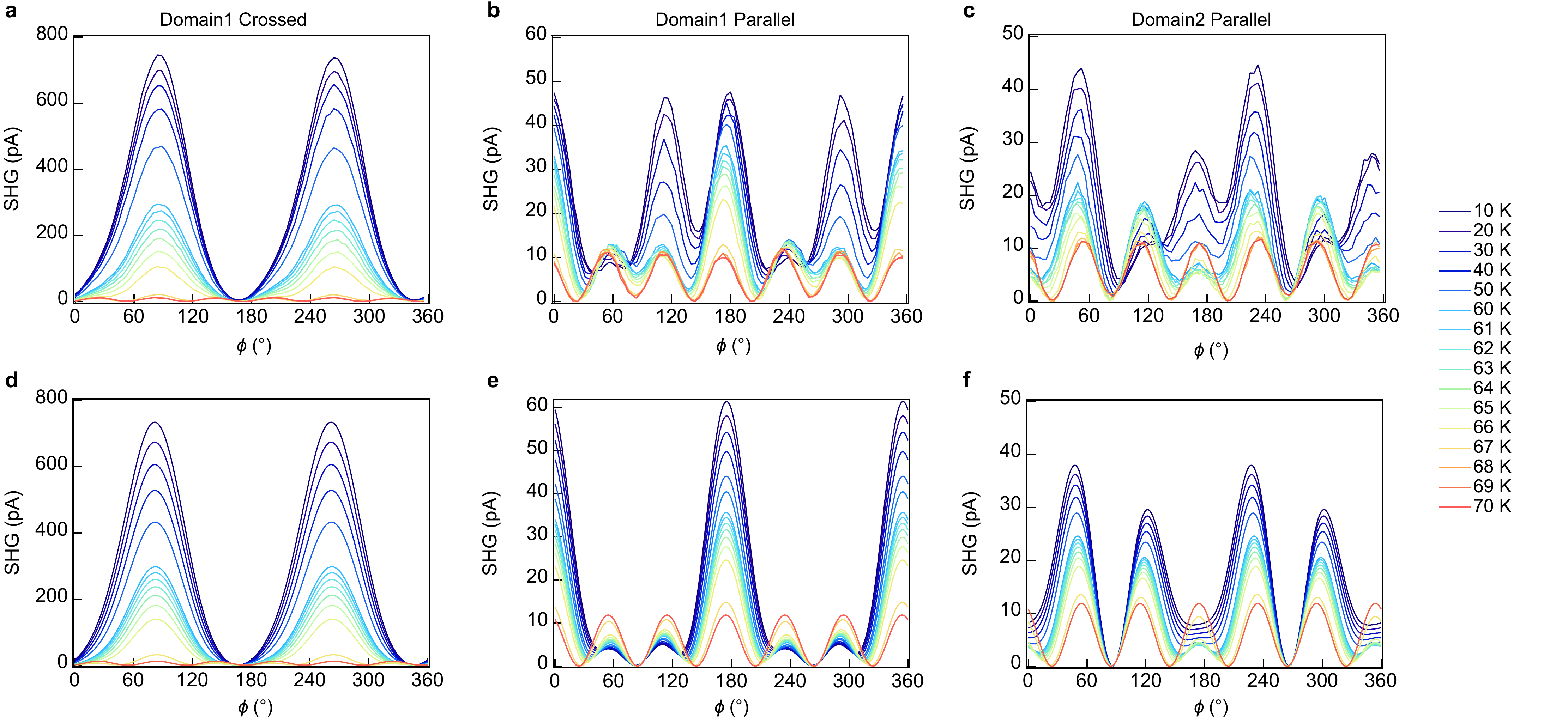}
\caption{\textbf{Temperature dependence of SHG patterns in an $\sim$ 50 nm thick MnPSe$_3$ flake on SiO$_2$/Si.} \textbf{a}, Crossed and \textbf{b} parallel patterns in domain 1. \textbf{c} Parallel patterns in domain 2. \textbf{d-f} Simulated crossed (\textbf{d)} and parallel (\textbf{e}) patterns in domain 1 and (\textbf{f)} parallel pattern in domain 2  by using Equa.\ref{couplingcrossed} with $L=(1-T/T_N)^\beta$. }
\label{tempdependence}
\end{figure}

\subsection{Temperature dependence of SHG patterns in a $\sim$50 nm thick flake exfoliated on SiO$_2$/Si}
Here we examine the validity of neglecting the higher-order terms in the Taylor expansion of second-order susceptibility. When the temperature is close enough to the N\'eel temperature, the N\'eel vector $\mathbf{L}$ is small, and therefore, it is reasonable to neglect higher-order terms of $\mathbf{L}$. However, when the sample temperature is far below the N\'eel temperature, all of the 6 non-zero terms in second-order susceptibility $\chi^{(2)}$ are necessary. To test this, we measure the temperature dependence of SHG patterns at one spot of a $\sim$ 50 nm thick flake. In Supplementary Figure \ref{tempdependence}a,b, we show the data for crossed and parallel patterns in domain 1 and in Supplementary Figure \ref{tempdependence}c, we show the parallel pattern for domain 2. Using Equa.\ref{couplingparallel} and \ref{couplingcrossed} we perform a simulation of the experiment data with $L=(1-T/T_N)^\beta$ (shown in Supplementary Figure \ref{tempdependence}d-f). The experiment and simulation match well across the whole temperature in crossed patterns. In parallel patterns, however, the simulation using Equa. \ref{couplingparallel} matches the experiment near N\'eel temperature: in both domains 1 and 2, the experimental data is reproduced by the simulation above 60 K, but not captured well below 50 K. We also test a 4th 100-nm sample and reached the sample conclusion that Equa. \ref{couplingparallel} works above 60 K. The break-down in parallel pattern indicates other $\chi^{ED}_{ijk}$ terms should be considered at low temperature, while for the crossed pattern, Equa. \ref{couplingcrossed} is good enough to capture the key factors such as the amplitude and the node direction.

\section{DFT calculation of the polar pattern from ED contribution}

To see how the polar pattern looks like by considering all of the non-zero terms, we performed density functional theory calculation. Electronic structure of antiferromagnetically ordered monolayer MnPSe$_3$ with spin polarization
aligned along $x$ direction ($a$ axis) shown in Supplementary Figure \ref{DFT}a. was calculated using first-principles density-functional theory (DFT) implemented in the Vienna \textit{ab-initio} Simulation Package with a plane-wave basis and the projector-augmented wave method. We adopted the Perdew-Burke-Ernzerhof (PBE)'s form of exchange-correlation functional within the generalized-gradient approximation (GGA) and a Monkhorst-Pack k-point sampling for the Brillouin zone (BZ) integration. An energy cutoff of 300 eV for the plane-wave basis and a Monkhorst-Pack k-point sampling of 24 $\times$ 24 $\times$ 1 were applied. Spin-orbit coupling was taken into account at the full-relativistic level. Hubbard $U$ correction was included in the DFT-PBE calculations with $U_{eff} =U-J=$ 3.9 eV to account for the correlation effect from $3d$ transition metal. The monolayer structure was extracted from the experimental bulk structure \cite{wiedenmannssc81} which holds in-plane antiferromagnetically-ordered monolayer MnPSe$_3$ with spin polarization aligned along $x$ direction (AFM-x, see Supplementary Figure \ref{DFT}a). The electronic band structure and spin density along x were shown in Supplementary Figure \ref{DFT}b. The calculated band-gap is $\sim$ 1.6 eV, which was smaller than the experiment value 2.32 eV measured on a bulk sample \cite{grassoJOSAB99} as expected for PBE+U. Our room temperature optical conductivity measurement shown in Figure \ref{linear} is qualitatively the same as the previous work \cite{grassoJOSAB99}.  Moreover, the spin density is localized around Mn atoms with spin polarization along $x$ axis. Note that the calculation does not take the exciton effect into consideration.

\begin{figure}
\centering
\includegraphics[width=0.7\textwidth]{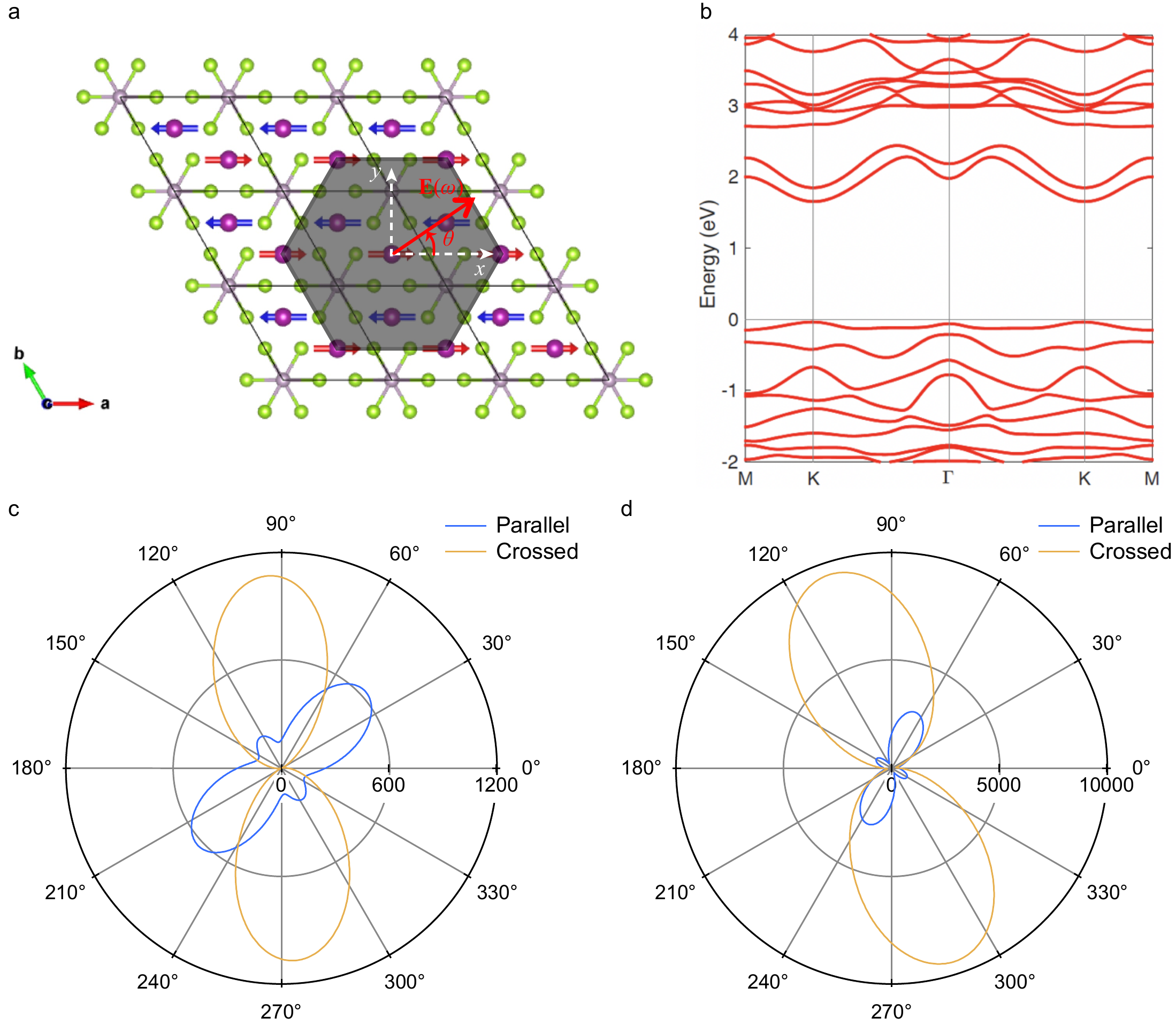}
\caption{\textbf{DFT results.} \textbf{a}, Crystal structure of monolayer MnPSe$_3$ with AFM-x magnetic ordering. \textbf{b}, Electronic band structure of AFM-x MnPSe$_3$. \textbf{c-d}, Calculated parallel (blue) and crossed (gold) polar patterns at the incident photon energy of (\textbf{c}), 0.83 eV and (\textbf{d}), 1.1 eV.}
\label{DFT}
\end{figure}

The magnetic point group of multi-layer (with the N\'eel vector along $a$ axis) is $\overline{1}' $; hence the parity-time-symmetry allows c-type SHG with all components being symmetry allowed. SHG susceptibility tensor was calculated by using an in-house developed first-principles nonlinear optics package (iNLO)\cite{huananolett2017,huasciadv2019} interfaced with VASP. Total 100 electronic bands were included in the calculations, and a small imaginary smearing factor of 0.05 eV was included in the fundamental frequency in the denominator of susceptibility tensor. The result at the incident photon energy of 0.83 eV is shown in Supplementary Figure \ref{DFT}c, considering that the gap size in the calculation is smaller by 0.73 eV and the incident photon energy in the experiment is centered at 1.55 eV. The anisotropy plot indicates that the lobes in the crossed pattern are mainly pointing perpendicular to the direction of the N\'eel vector. In other words, the node direction in the crossed pattern is close to the N\'eel vector direction, which supports the above analysis in Supplemantary Note 1. We also point out that we looked at the node direction over a wide energy range from 0.6 eV to 1.2 eV for the incident photon energy in the calculation. It is always within $\sim$ 30$^\circ$ from the N\'eel vector. An example of the polar patterns at 1.1 eV is shown in Supplementary Figure \ref{DFT}d, which also looks quite similar to the experiment.

\section{More SHG data}
\subsection{More SHG imaging about AFM domains on a bulk crystal}
Here we show SHG mapping about AFM domains on a $\sim$ 30 $\mu$m bulk crystal in Supplementary Figure \ref{mapping}. The incident and detecting light polarization are shown on the top of each figure. Supplementary Figure \ref{mapping}a. is measured when the peak of the crossed pattern is reached. The darker lines indicated by the white arrows are scratches on the sample surface, while other darker lines indicate the AFM domain walls between 180$^\circ$ domains. Supplementary Figure \ref{mapping}b,c. show SHG intensity mapping in the parallel channel with two different polarizer angles. Two different domains are clearly distinguishable by different SHG intensity in both mappings with sharp contrasts. The outlines of the domains in the three figures match well. These observations are consistent with our previous discussion on the domain wall between two Ising domains.

\begin{figure}
\centering
\includegraphics[width=0.7\textwidth]{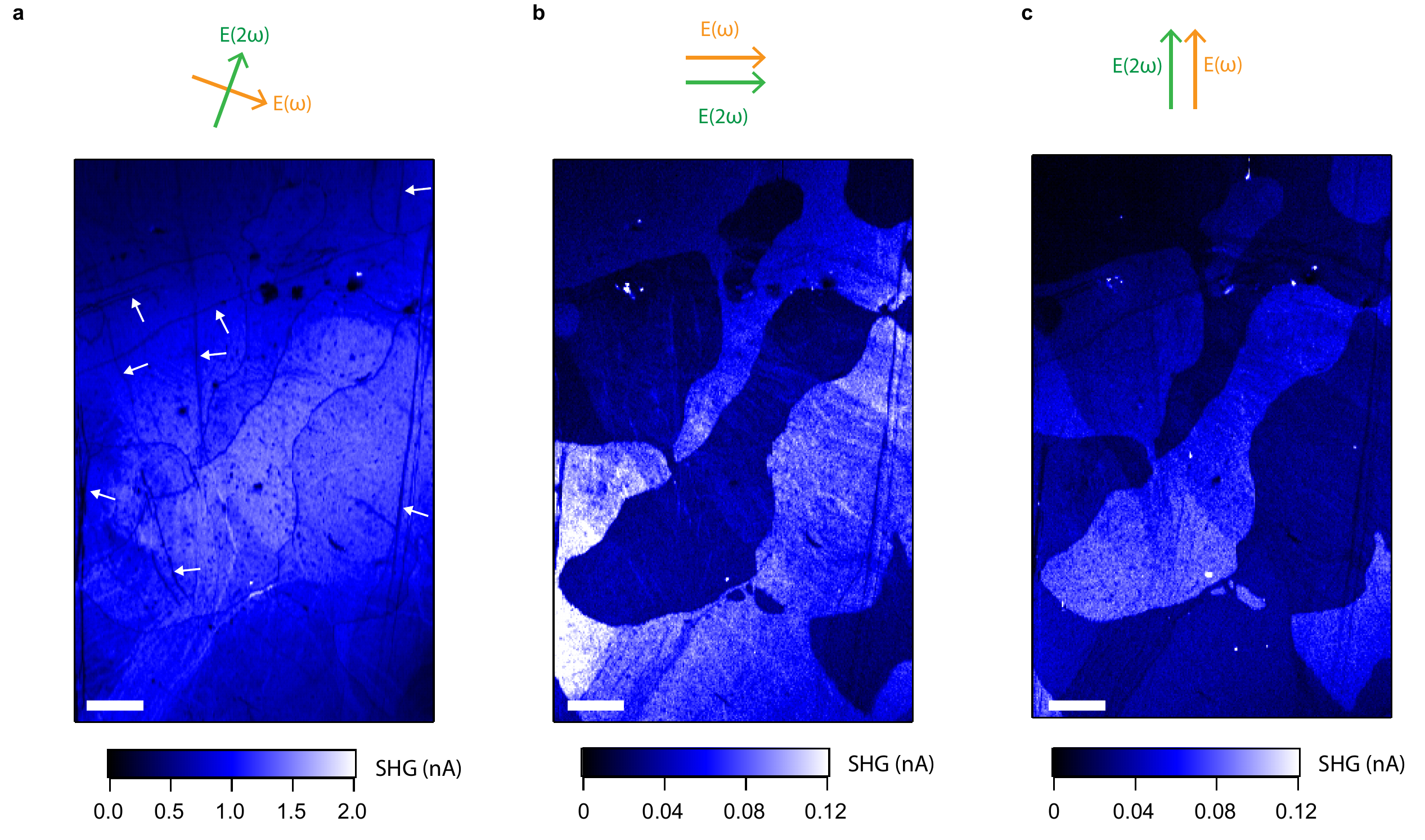}
\caption{\textbf{SHG intensity mapping in crossed and parallel configurations of a MnPSe$_3$ bulk crystal ($\sim$30 $\mu$m thick) at 5 K.} \textbf{a}, Mapping with polarizers at the peak of the crossed pattern. The dark lines marked by white arrows are scratches on the sample surface. The dark lines without white arrows are mobile AFM domain walls. \textbf{b}, Mapping with polarizers at the peak of the parallel pattern in one of the domains. \textbf{c}, Mapping with polarizers at the peak of the parallel pattern of the other domain. Scale bar: 50 $\mu$m. The incident and detecting polarization are marked by orange and green arrows, respectively. }
\label{mapping}
\end{figure}

\subsection{SHG polar patterns and mapping on atomically thin few-layer MnPSe$_3$}

Supplementary Figure \ref{multilayermapping} shows an SHG intensity mapping in a multi-layer sample with different domains. The measuring polarization is chosen to be near the peak of areas marked by the red arrow. Regions with homogeneous thickness and same AFM domains are outlined by white lines (and gray line for the 4-layer sample) . The arrows indicate the node direction in the crossed patterns. In each region, the crossed polar pattern is measured and shown beside the intensity mapping. Note that part of the 2L sample with the tail shape on the bottom left corner has a different N\'eel vector direction probably due to a different strain direction. Nevertheless, as we mentioned in the main text, all of the samples measured in this work, including these few-layer ones, all show two-state (Ising) ground state with in-plane spins across thermal cycles at randomly selected positions we measure.

\begin{figure}
\centering
\includegraphics[width=0.6\textwidth]{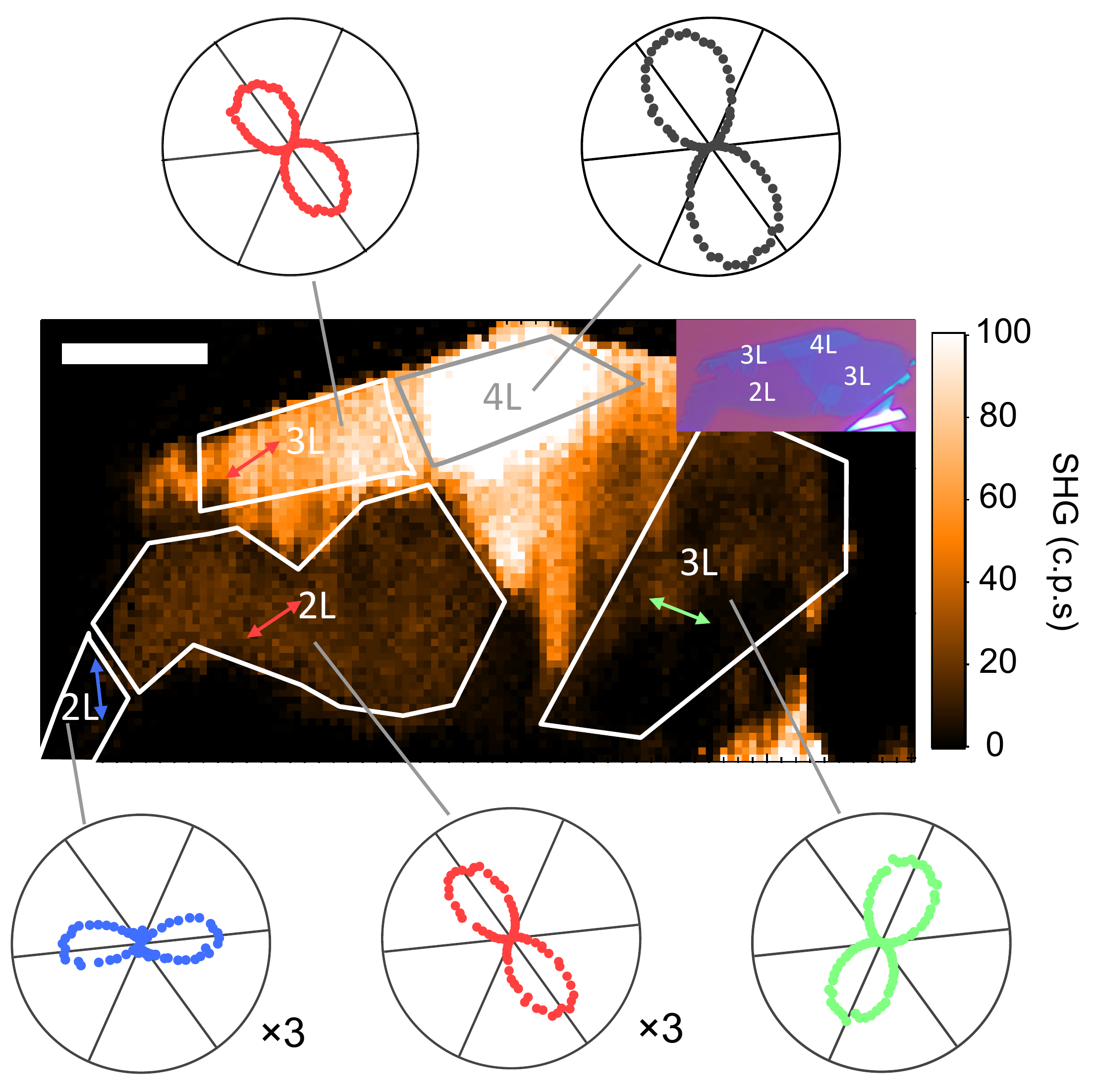}
\caption{\textbf{N\'eel vector mapping of an atomically thin multilayer MnPSe$_3$ sample  exfoliated on the SiO$_2$/Si substrate in air.} SHG intensity mapping of a MnPSe$_3$ flake with various layers. Areas with the same thickness and N\'eel vector direction are noted by the solid lines and labeled with the layer number. An optical image of the same flake is shown in the corner. The two-way arrows in different colors denote the node orientation (N\'eel vector direction) of the crossed pattern in the marked area. The measurement is performed at 5 K.  Scale bar: 10 $\mu$m. }
\label{multilayermapping}
\end{figure}

\subsection{Critical behavior fitting}
The order parameter follows $(1-T/T_N)^\beta$ with $\beta$ being the critical exponent in a continuous phase transition. Since $\chi^{ED}$ is proportional to the order parameter \cite{savalenti00}, on can get $I_{SHG}^{1/2}\propto(1-T/T_N)^{\beta}+b$ where $b$ is a constant coming from the EQ contribution. In Supplementary Figure \ref{betafitting}, we plot $(I_{SHG}^{1/2}-b)$ that was shown in Extended Data Figure 5 as a function as $(T_N-T)/T_N$ in a log-log plot. The slope of the linear curve near the origin is the critical exponent $\beta$. The experiment data and the best fit of bulk (a), trilayer (b), bilayer (c) and monolayer (d) samples are showed. Note the fitting error is pretty high in monolayer due to the uncertainty of the N\'eel temperature. Overall, the critical exponent is $\sim$ 0.3, and the critical exponential of 0.29 ($\pm$ 0.01) is in the 15 $\mu$m thick bulk crystal is most reliable considering the very dense temperature step in the measurement and the highest signal-to-noise ratio. From the critical component itself, even though it is close to the 3D Ising model ($\beta=0.3265$) than the 3D XY model ($\beta=0.348$), it is not very obvious that the phase transition belongs to the Ising universality class since the difference in $\beta$ is small. Nevertheless, the collapse of the temperature-dependent SHG on two curves is a strong evidence for the Ising order.

\begin{figure}

\centering
\includegraphics[width=0.6\textwidth]{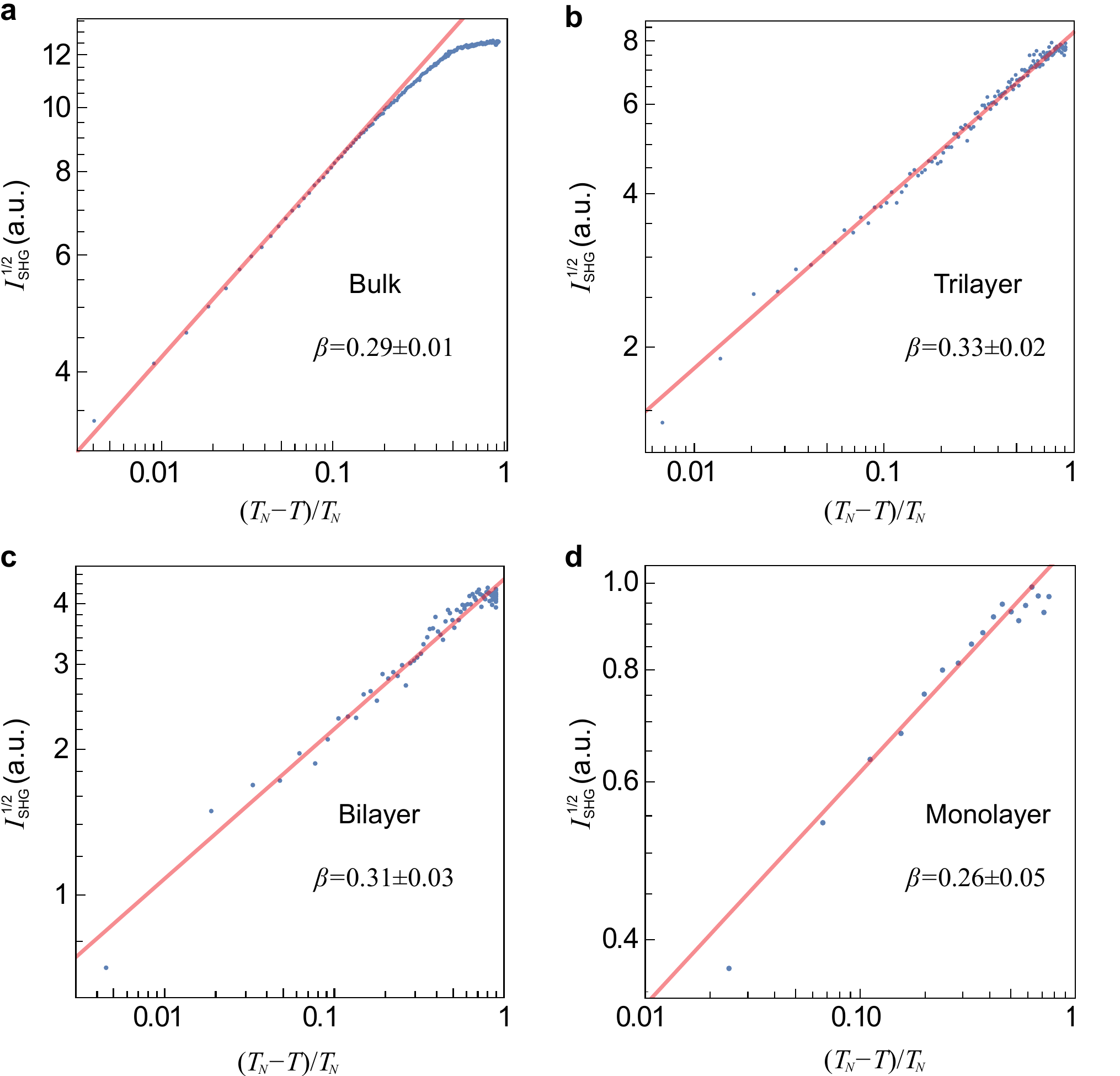}
\caption{\textbf{Critcal behaviors of  MnPSe$_3$ samples shown in extended data Fig. 5}. The temperature dependence of SHG intensity is shown in log-log plot. A linear fit to the data is applied to the linear region of each curves. \textbf{a}, 15 $\mu$m thick bulk. \textbf{b}, Trilayer. \textbf{c}, Bilayer. \textbf{d}, Monolayer. Few-layer samples in \textbf{b-d} are exfoliated in air.}
\label{betafitting}
\end{figure}

\subsection{N\'eel vector distribution in multi-layer MnPSe$_3$}
Since the crossed pattern nodes are locked to the N\'eel vectors, it is possible to map the spatial distribution of N\'eel vector within a MnPSe$_{3}$ flake. 
To perform N\'eel vector mapping on thick flakes of MnPSe$_3$, a flat area of a 80 $\mu$m $\times$ 80 $\mu$m on a 100-nm MnPSe$_3$ flake is chosen as indicated by the black square in Supplementary Figure \ref{bulkmapping}a. Contrary to the uniform image under an optical microscope (Supplementary Figure \ref{bulkmapping}a), the SHG signal, however, is highly inhomogeneous in this area (Supplementary Figure \ref{bulkmapping}b). First, we measure the crossed patterns in the selected points marked with white circles in Supplementary Figure \ref{bulkmapping}b. Both the magnitude and the orientation of N\'eel vectors are non-uniform revealed by crossed patterns shown in Supplementary Figure \ref{bulkmapping}c. To map the crossed patterns in the whole region, we measured the crossed polar patterns in each spot with a 2 $\mu$m size and fit them by $L^2\sin^2{(\theta-\phi)}$ to get the direction of the N\'eel vectors first. The lines in \ref{bulkmapping}d represent the spatial distribution of the extracted directions of the N\'eel vectors. The directions of the node in crossed patterns are shown by the orientation of the line as well as the color of the line segments. Supplementary Figure \ref{bulkmapping}e shows the histogram of the orientations of N\'eel vectors.

\begin{figure}
\centering
\includegraphics[width=0.8\textwidth]{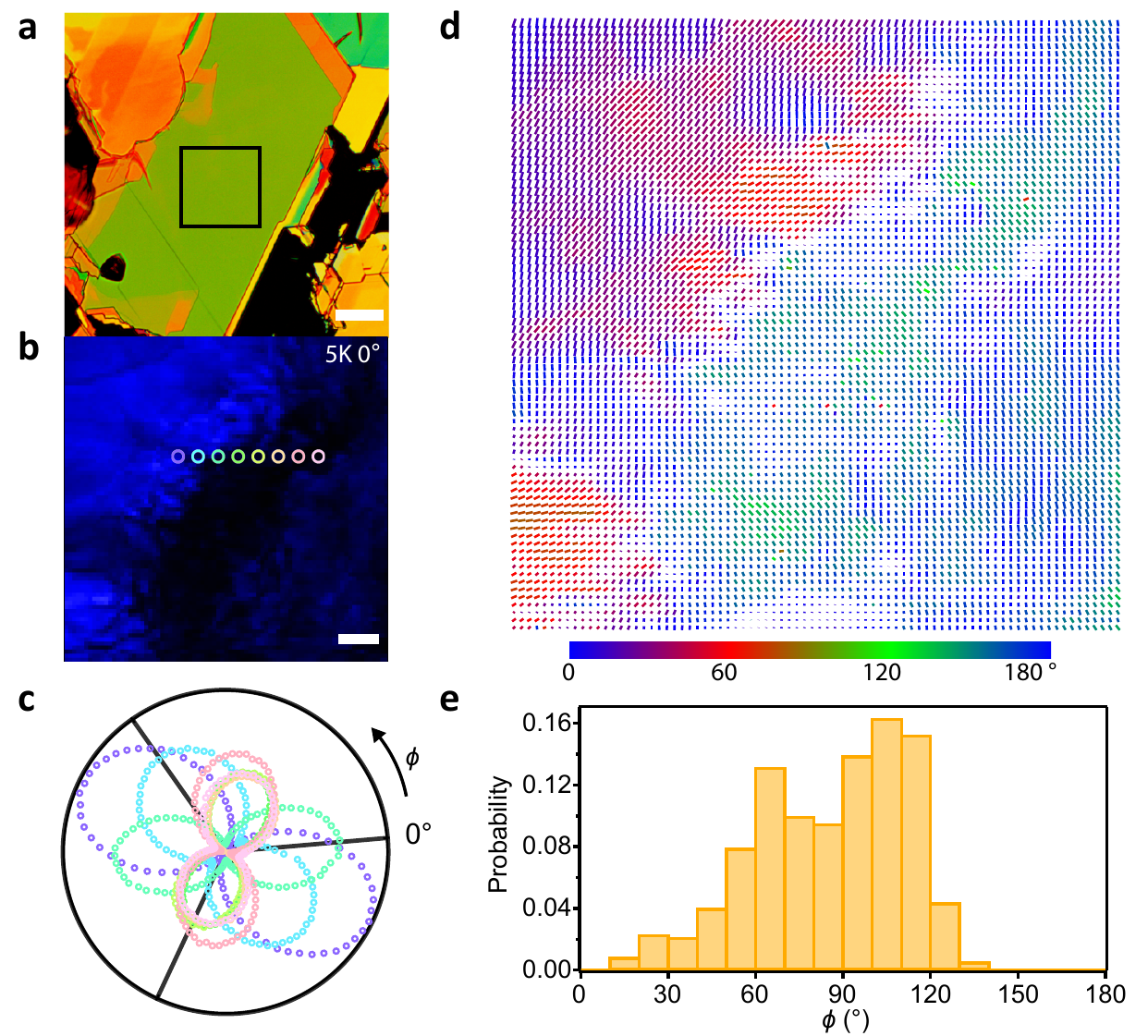}
\caption{\textbf{Distribution of N\'eel vectors in a (fourth) as-exfoliated 100 nm thick MnPSe$_3$ flake on the SiO$_2$/Si substrate. } {\bf a}, Optical image of the flake . The green area in the center has a homogeneous thickness of around 100 nm. Scale bar: 50 $\mu$m. {\bf b}, SHG intensity image of the squared area in {\bf a} with polarizers at 0$^\circ$ of the crossed pattern. The measured area is 80 $\mu$m$ \times $80 $\mu$m. Scale bar: 10 $\mu$m {\bf c}, Crossed patterns in varying points. The position where patterns are taken are marked in {\bf b}. {\bf d}, Vector mapping of N\'eel vectors in the same area as {\bf b}. The  orientation of each line segment denotes the  direction of the N\'eel vectors. Different orientations are marked by different colors. {\bf e}, Histogram of orientations in {\bf d}. The measurement is performed at 5 K.}
\label{bulkmapping}
\end{figure}

According to our previous analysis, the magnitude of the crossed pattern should not change if the direction of N\'eel vector changes in an unstrained thick flake. However, the $L^2$ from the fit shows the magnitude of crossed patterns is usually not the same at different positions. The magnitude of the N\'eel vector is represented by the length of the line segments. To resolve this inconsistency, we hypothesize that the orientation of the N\'eel vector could be different in different layers, probably due to the weak interlayer coupling.

We consider a multi-layer sample consisting of a few layers in a domain where the node direction of the crossed pattern points to $\theta=0^\circ$ and another few layers with the node direction pointed along $\theta=120^\circ$. For simplify we assume each layer produces same second-harmonic electric field. Then we can derive the total SHG signal in crossed pattern detected with incident polarization $\phi$ to be
\begin{align}
I^{total}_{crossed}
&\propto\left(A\sin{\Delta\theta+B\sin{(\Delta\theta-\frac{2\pi}{3})}}\right)^2\\
&\propto\left(\alpha\sin{(\Delta\theta+\gamma)}\right)^2,
\end{align}
where $\Delta\theta=\phi-\theta$ and $\alpha$ and $\gamma$ are constants determined by A and B. By tuning A and B, we can observe a crossed pattern with an arbitrary direction and amplitude for the N\'eel vector. Therefore, the experiment results in unstrained thick MnPSe$_3$ could be explained by contributions between different N\'eel vectors in different layers.

\subsection{Aging effect of a bilayer MnPSe$_3$ sample in air}

We tested the degradation of a bilayer MnPSe$_3$ by exposing it to air. We exposed the same sample in air for 20 mins, 4 hours, and 20 hours and measured the SHG intensity and the transition temperature. During each exposure, the sample was stored in an atmospheric environment with relative humidity less than 30\%. With the increase of exposure time, the SHG intensity and the N\'eel temperature decrease (Supplementary Figure \ref{degradation}). We conclude that the MnPSe$_3$ is air-sensitive in its ultra-thin forms.

\begin{figure}
\centering
\includegraphics[width=0.5\textwidth]{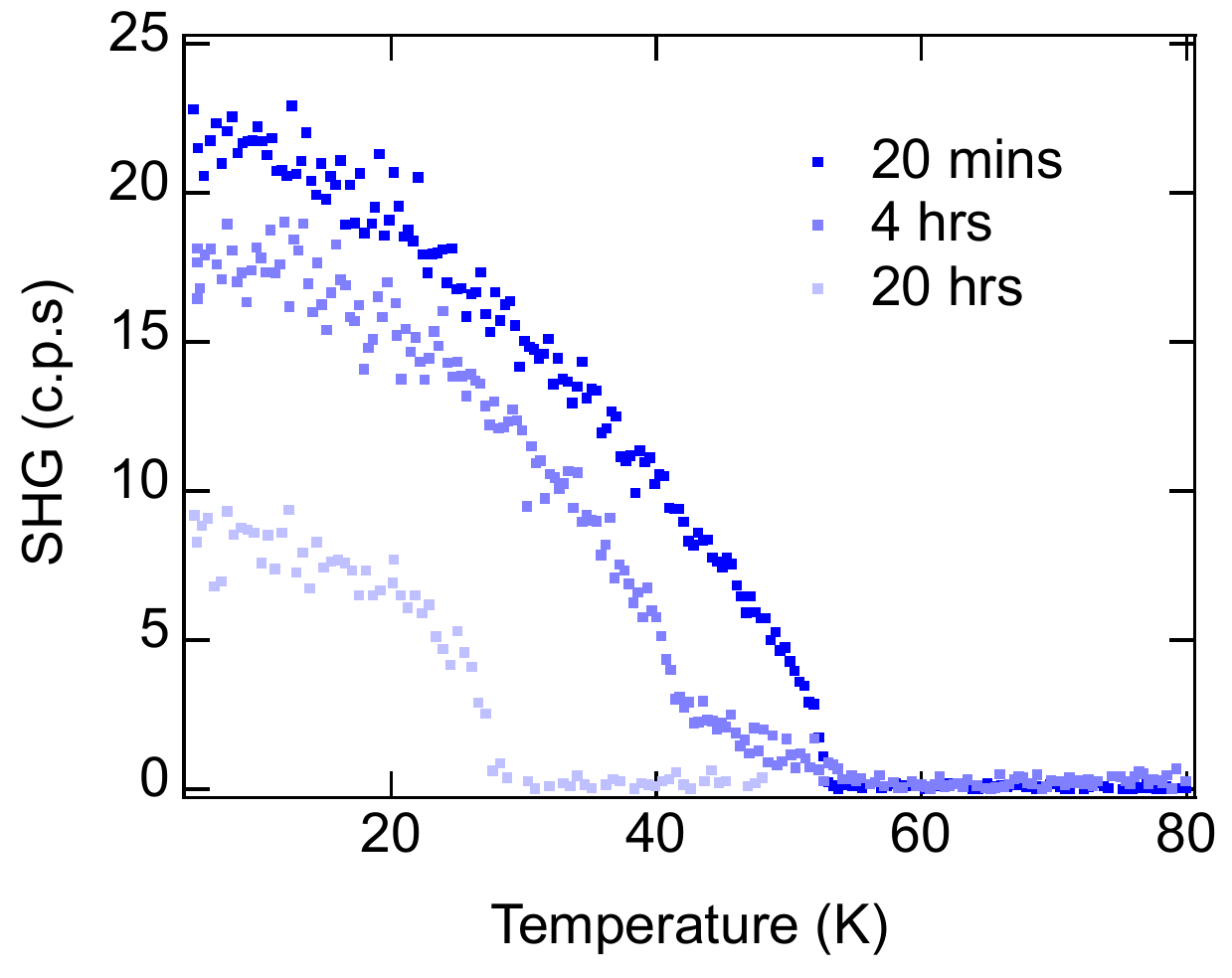}
\caption{\textbf{Aging effect of an as-exfoliated bilayer MnPSe$_{3}$ sample in air}.}
\label{degradation}
\end{figure}

\begin{figure}
\centering
\includegraphics[width=0.5\textwidth]{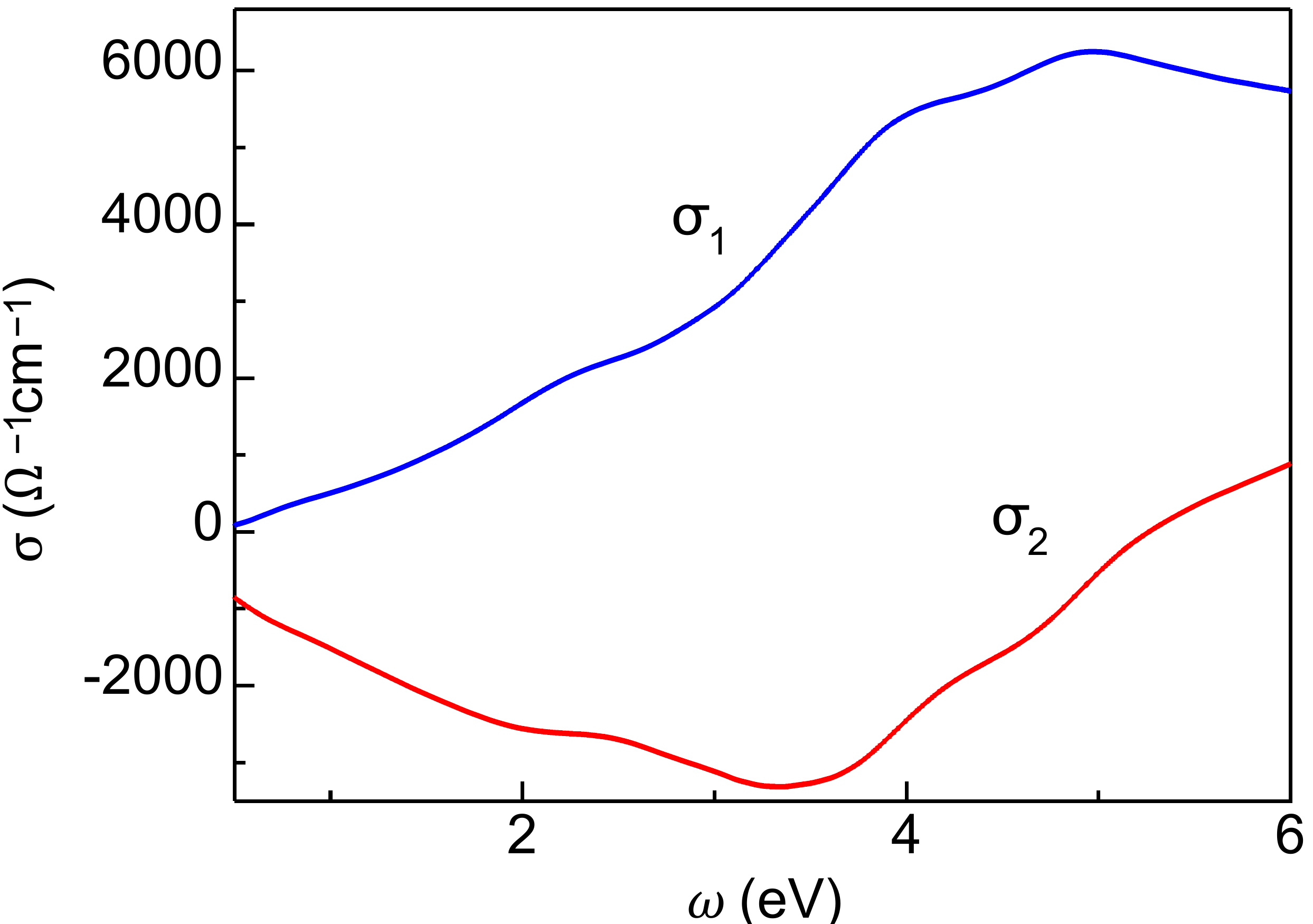}
\caption{\textbf{Optical conductivity of a MnPSe$_3$ bulk crystal at room temperature.}}
\label{linear}
\end{figure}

\section{Strain dependent N\'eel vector distribution}

First, we discuss how we determine the strain strength in MnPSe$_3$ flake and PDMS substrates. We show optical images of the strained sample presented in Fig. 4 in the main text. Images on the sample as-exfoliated and under uniaxial strain in four different directions are shown in Supplementary Figure \ref{strainimage}. In each direction, a $\sim 15\%$ strain is added to the PDMS substrate by a micro-manipulator. Supplementary Figure \ref{strainimage}a and b show the distance change between two bulk flakes (marked by the solid lines) before and after applying the strain, where elongation of 15-16\% is observed in the strained substrate. We also made an estimation that around 2\% strain is added to the sample by comparing the change of the length of the sample before and after adding the strain (shown in Supplementary Figure \ref{strainimage}c-e). Note that the error bar is around $\pm$ 0.5 $\%$, and the strain strength on different sample positions is possible to be slightly different because of the finite sample size \cite{liunatcomm14}. In Supplementary Figure \ref{strainimage}d-g, optical images of the sample with different strain directions are shown. Nevertheless, as  Fig. 4 in the main text shows that the SHG polar patterns are  rotated and that the SHG intensity is uniform, strain is obviously added to the sample to align the N\'eel vectors. We hope that future Raman scattering experiments could determine the strain direction and strength more accurately. In our experimental, we could control strain direction within 10 degree, and the direction is more important. As shown in the microscopic spin model,  small strain locks the N\'eel vector direction, and increasing the strain strength does not further tip the N\'eel vector in the linear response regime. Our SHG polar pattern measurement shows that the N\'eel vector is aligned towards the strain direction, and therefore, the conclusion of strain-controlled Ising order is valid even though the method of determining the strain strength by optical images (with cracks in the samples sometimes) might be less accurate than Raman scattering.

\begin{figure}
\centering
\includegraphics[width=1\textwidth]{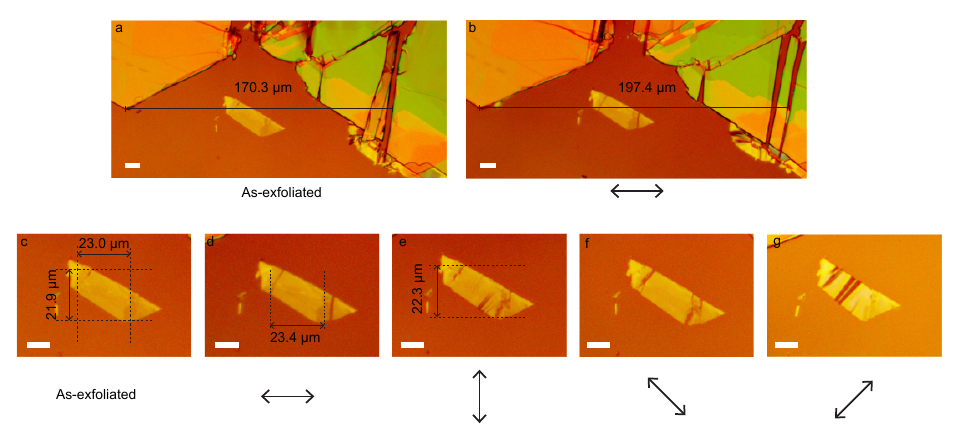}
\caption{\textbf{Optical images of the MnPSe$_3$ sample  in Fig.4 of the main text under strain.} \textbf{a-b}, Optical image of the MnPSe$_3$ flake ($\sim$ 15 nm thick) on a PDMS substrate without being stretched (\textbf{a}) and a horizontally strained PDMS substrate (\textbf{b}). An elongation of 15-16\% is observed by measuring the distance (solid line) between two bulk flakes. \textbf{c-g}, Optical images of the MnPSe$_3$ sample under strain in different directions. The two-way arrows beneath each image indicate the strain direction. A uniaxial $\sim$ 15\% strain is added to the PDMS substrate. A roughly $\sim$ 2\% strain is transferred onto the sample, which is estimated by measuring the sample length change in the optical images. Scale bar: 10 $\mu$m.}
\label{strainimage}
\end{figure}

We investigate the effect of the strain on the N\'eel vector by applying different strength of strain along one direction on a MnPSe$_3$ flake with a thickness of 10 nm determined by atomic force microscopy measurement. The data are shown in Supplementary Figure \ref{strainmapping}. 7.5\% and 15\% strain are added to the PMDS substrate and the strain on the sample is estimated by the strain on the PMDS and a  transfer ratio of 13\%. When the sample is as-exfoliated and under 1\% and 2\% elongation, we map the node direction in the crossed patterns using the same method as described in Equa. \ref{couplingcrossed}. When the sample is as-exfoliated, the node directions do not point to a specific direction. When 1\% strain is applied, the node direction in most regions points to the direction of the stretched direction except the region on the right edge. When we further add strain to 2\%, the node direction points to the stretched direction across the whole sample. Note that because of the small sample size, the strain direction near the edge might be more influenced by the sample shape or  the substrate during exfoliation process instead of the externally applied strain, which may explain the observation that it is harder to align the  N\'eel vectors on the edges.

\begin{figure}
\centering
\includegraphics[width=0.8\textwidth]{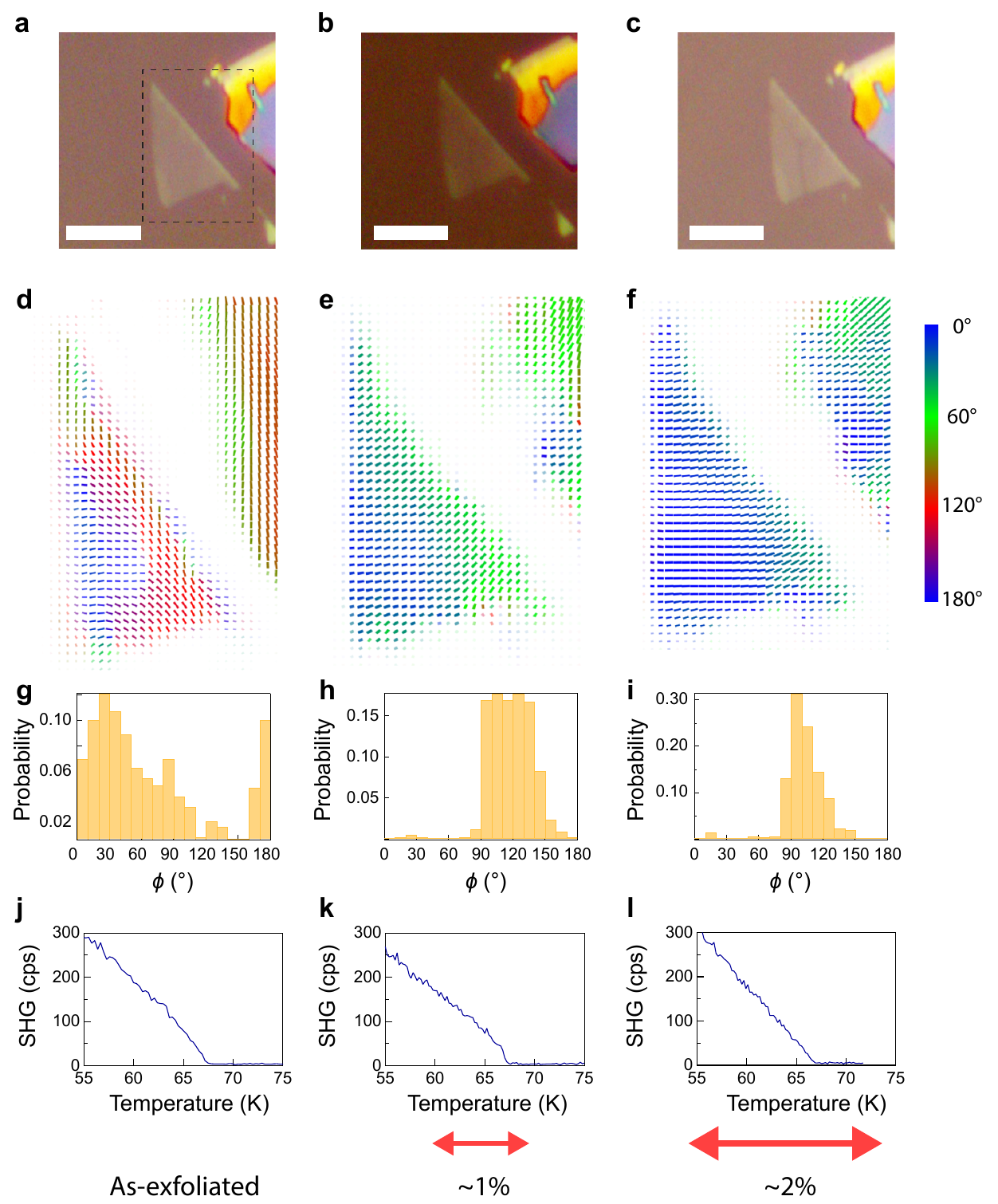}
\caption{\textbf{N\'eel vector distribution in a 10-nm thick MnPSe$_3$ flake on a PDMS substrate under different strain strength.} \textbf{a-c}, Optical image as-exfoliated, under 1\% strain and 2\% strain. The strain strength is estimated by the strain added to the PDMS times an estimated strain transfer rate  of 13\%. The sample thickness is measured by the atomic force microscopy. Scale bar: 10 $\mu$m. \textbf{d-f}, N\'eel vector (crossed pattern node) direction mapping at 5 K when the sample is as-exfoliated and under 1\% strain and 2\% strain. \textbf{g-i}, Histogram of the N\'eel 
vector (crossed pattern node) direction when the sample is as-exfoliated and under 1\% strain and 2\% strain. \textbf{j-l}, SHG intensity as a function of temperature when the sample is as-exfoliated and under 1\% strain and 2\% strain.}
\label{strainmapping}
\end{figure}

The PDMS substrate is not a good thermal conductor when it is 30 $\mu$m thick, and it can induce a $\sim$10 K temperature difference between the sample and the platform, which is calibrated by measuring the transition temperature of a thick ($>$ 50 nm) flake sample on the same PDMS substrate. We then measure the $T_N$ with calibrated temperature as shown in Supplementary Figure \ref{strainmapping}j-i. There is no observable $T_N$ change (within 1 K) between these three cases.

\section {Landau theory}

Here we consider a Landau theory for the staggered magnetization $M_i$. 
Symmetries constrain the terms that can appear in the expansion.   We have the rotationally invariant  tensor $\delta_{ij}$, which allows $|\vec M|^2$, and planar symmetry allows $M_z^2$.   We assume that this term is large, so that $M_z=0$.   For $M$ in the $xy$ plane, the 3-fold rotation symmetry, along with time reversal allows a term $(M_x + i M_y)^6 + c.c.$.  This identifies an allowed rank 6 tensor that can be contracted with $M_i$.   

Strain is described by a symmetric second rank tensor $u_{ij}$, whose principal axes describe the directions of compressive and tensile strain.   To lowest order in the strain, a term $u_{ij} M_i M_j$ is allowed.

It is easiest to describe the magnetization in polar coordinates, $M_x + i M_y = M e^{i \theta}$.   Then, the crystalline anisotropy gives $M^6 \cos 6\theta$.   The symmetric strain can be diagonalized with a rotation about the $c$ axis by an angle $\theta_0$.   This term then gives $u_s M^2 \cos 2(\theta-\theta_0)$.

The Landau expansion (to lowest order in the coupling to strain) then takes the form
\begin{equation}
F = - a t M^2 + b u_s M^2 \cos 2(\theta-\theta_0) +c M^4 + d M^6 + e M^6 \cos 6\theta
\end{equation}
where $t = (T^0_c-T)/T^0_c$.  Here $T_c^0$ is the mean field critical temperature for $u_s=0$. $M=L/2$, where $L$ is the magnitude of the N\'eel vector.

Minimizing with respect to $M$ and $\theta$ is equivalent to doing mean field theory.   For $0<t\ll 1$, this gives 
\begin{equation}
M_0 = (at/2c)^{1/2} 
\end{equation}
with the mean field exponent $\beta=1/2$.  

For the critical behavior fluctuations are important and mean field theory breaks down.  To get a better description of this, we can integrate out massive fluctuations in the magnitude $M$ to obtain an effective theory for $\theta$, which has the form,
\begin{equation}
F(\theta) =  b u_s M_0^2 \cos 2(\theta - \theta_0) + e M_0^6 \cos 6\theta
\end{equation}
As shown in main text Fig. 5, for $u_s=0$ this reduces to the 6 state clock model.  For $u_s \ne 0$ it is in the Ising universality class, and the antiferromagnetic order parameter is locked to the strain axis with a constant tilt angle.

\section {Spin model}

Here we construct a more microscopic spin model to explain the spin-strain locking. Space group 148 is a chiral group with the point symmetry $C_3$. The high temperature state has two Mn sites on a honeycomb lattice. Below the N$\rm{\acute{e}}$el temperature the two sublattice sites develop different spin polarizations. Thus the AF transition is a condensation of two ${\mathbf{G}}=0$ fields giving the spin polarizations on the $\alpha$ and $\beta$ sublattice sites. Notice that this does not lower the translational symmetry of the structure, but only assigns  the spins to the two sublattice sites.

\smallskip
The starting model for the structure has $xy$ easy plane anisotropy. So the minimal model has on site anisotropy
\begin{eqnarray}
{\cal H} = J \sum_{{\rm nn}: \,  \langle i \in \alpha, j \in \beta \rangle }  \, {\mathbf{s}}_i \cdot {\mathbf{s}}_j + J_z \sum_{\forall i} \left( {\mathbf{s}}_{i} \cdot \hat z \right)^2 \nonumber
\end{eqnarray}
where $J>0$ and $J_z > 0$.

\smallskip
Strain is a time reversal even, traceless second rank tensor $\varepsilon_{\mu \nu}$. Each link of the honeycomb lattice can be defined by an in-plane vector ${\mathbf{d}}_{ij} = {\mathbf{r}}_{(j \in \beta)} - {\mathbf{r}}_{(i \in \alpha)}$ and in the presence of strain we deform the links according to
\begin{eqnarray}
{\mathbf{d}}_{ij} \mapsto {\mathbf{d}}_{ij} + {\boldsymbol{\varepsilon}} \cdot {\mathbf{d}}_{ij} \nonumber
\end{eqnarray}
In the following we are going to treat just the traceless part of ${\boldsymbol{\varepsilon}}$  which allows for deformation but not dilation or compression. The change of the exchange constants in any bond should be $\propto \varepsilon_{\mu \nu}$ but {\it even} under ${\mathbf{d}}_{ij} \mapsto - {\mathbf{d}}_{ij}$. We build the simplest model that does this.  For any link define the unit vector $\hat {\mathbf{d}}_{ij}$ and the projections
\begin{eqnarray}
{\mathbf{s}}_i \cdot \hat {\mathbf{d}}_{ij}\,  &,& \,  {\mathbf{s}}_i - \left({\mathbf{s}}_i \cdot \hat {\mathbf{d}}_{ij} \right) \hat {\mathbf{d}}_{ij}  \nonumber\\
{\mathbf{s}}_j \cdot \hat {\mathbf{d}}_{ij} \,&,& \,  {\mathbf{s}}_j - \left({\mathbf{s}}_j \cdot \hat {\mathbf{d}}_{ij} \right) \hat {\mathbf{d}}_{ij} \nonumber
\end{eqnarray}
To lock the spin direction to the strain we have to replace the isotropic $xy$ coupling in the original model by
\begin{eqnarray}
J \, {\mathbf{s}}_i \cdot {\mathbf{s}}_j \mapsto J_{\|,ij} {\mathbf{s}}_{i, \|}  {\mathbf{s}}_{j, \|} + J_{\perp,ij} {\mathbf{s}}_{i, \perp}  {\mathbf{s}}_{j, \perp} + J_c  \left({\mathbf{s}}_{i, \|} {\mathbf{s}}_{j, \perp} + {\mathbf{s}}_{i, \perp} {\mathbf{s}}_{j, \|} \right)
\end{eqnarray}
distinguishing spin polarizations ``along" and ``perpendicular to" the $ij$-th link. If we keep the isotropic form of the nearest neighbor coupling it would exclude the possibility of any locking of the spin to the strain.  

Combining the first two terms gives an expression like
\begin{eqnarray}
J_\perp {\mathbf{s}}_i \cdot {\mathbf{s}}_j + \left( J_\| - J_{\perp} \right) \left( {\mathbf{s}}_i \cdot \hat {\mathbf{d}}_{ij} \right) \left( {\mathbf{s}}_j \cdot \hat {\mathbf{d}}_{ij} \right)\nonumber
\end{eqnarray}
so the first term just looks like the isotropic $xy$ coupling and the spin-lattice locking comes from the second piece. To make this more transparent we notice that $i \in \alpha$ and $j \in \beta$,  so in terms of the macroscopic spin fields ${\mathbf{S}}_\alpha$ and ${\mathbf{S}}_\beta$, it expresses as 
\begin{eqnarray}
J_\perp \, {\mathbf{S}}_\alpha \cdot {\mathbf{S}}_\beta +  \, {\mathbf{S}}_\alpha \cdot \left[ \sum_{{\rm nn}: \, {ij}} \, \Delta J_{ij}  \, {\mathbf{d}}_{ij} \otimes {\mathbf{d}}_{ij} \right] \, \cdot  {\mathbf{S}}_\beta \nonumber
\end{eqnarray}
where $\otimes$ denotes the outer product, i.e. in component form $\left[ \hat {\mathbf{d}} \otimes \hat {\mathbf{d}} \right]_{mn} = \hat d_m \hat d_n$. This contains an isotropic component  that can be removed by explicitly writing this as an outer product
\begin{eqnarray}
\left[ \hat {\mathbf{d}} \otimes \hat {\mathbf{d}} \right]_{mn} = \left( \hat d_m \hat d_n - \frac{1}{2} \,\,  \delta_{mn} \right) + \frac{1}{2} \,\, \delta_{mn} \nonumber
\end{eqnarray}
This means that one must distinguish between the ``longitudinal" and ``transverse" spin couplings in any  link but that the lattice sum produces an effective isotropic $xy$ exchange constant $J = (J_\| + J_\perp)/2$. The previous Landau expansion makes it clear that in the absence of strain the crystal field anisotropy can not appear at bilinear order in the spin Hamiltonian as indeed we find in this explicit model.

\smallskip
The above conclusion  demonstrates that if the $\Delta J$'s (which are symmetry allowed and therefore always present) were actually the {\it same} in each link then the traceless term in parenthesis gives zero after the sum over bonds because it is an $m=2$ tensor averaged over a crystal with threefold rotational symmetry.  Similarly, the lattice sum over the $J_c$ term gives zero since the prefactors have the symmetry of  an $m=2$ tensor.  However the trailing isotropic term gives a nonzero result after the lattice sum: $(J_\| - J_\perp) \,\, {\mathbf{S}}_\alpha \cdot {\mathbf{S}}_\beta/2$ and this contributes to the  isotropic $(xy)$ term of the spin Hamiltonian which then becomes
\begin{eqnarray}
J \, {\mathbf{S}}_\alpha \cdot {\mathbf{S}}_\beta =  \frac{J_\perp + J_\| }{2} \, \, {\mathbf{S}}_\alpha \cdot {\mathbf{S}}_\beta  \nonumber
\end{eqnarray}

\smallskip
\smallskip
The situation for the cross terms is similar.  We can write the symmetric cross term in the exchange in the form
\begin{eqnarray}
J_c \,  [({\mathbf{s}}_i \cdot \hat {\mathbf{d}})( \hat z \cdot (\hat{\mathbf{d}} \times {\mathbf{s}}_j)) &+&
( \hat z \cdot (\hat{\mathbf{d}} \times {\mathbf{s}}_i)) ( \hat {\mathbf{d}} \cdot {\mathbf{s}}_j )]  =  \nonumber\\
J_c \, [ ({\mathbf{s}}_i \cdot \hat {\mathbf{d}})( \hat z \times \hat{\mathbf{d}}) \cdot  {\mathbf{s}}_j) &+& ({\mathbf{s}}_i \cdot (\hat z \times \hat{\mathbf{d}}))( \hat {\mathbf{d}} \cdot {\mathbf{s}}_j)]  \nonumber
\end{eqnarray}
Using the notation $ \hat {\mathbf{d}} = (\hat d_x, \hat d_y)$  it is useful to write this
\begin{eqnarray}
J_c  \, {\mathbf{s}}_i \cdot \left[ \hat {\mathbf{d}} \otimes (\hat z \times \hat {\mathbf{d}}) + (\hat z \times \hat {\mathbf{d}}) \otimes \hat {\mathbf{d}} \right] \cdot {\mathbf{s}}_j \nonumber
\end{eqnarray}
Explicitly, the quantity in brackets is
\begin{eqnarray}
\left[ ... \right] =  2 \left[ - \hat d_x \hat d_y \left(
                                        \begin{array}{cc}
                                          1 & 0 \\
                                          0 & -1 \\
                                        \end{array}
                                      \right) + \frac{(d_x^2 - d_y^2)}{2} \left(
                                                                  \begin{array}{cc}
                                                                    0 & 1 \\
                                                                    1 & 0 \\
                                                                  \end{array}
                                                                \right)   \right] \nonumber
\end{eqnarray}
Because of the tensor character of the coefficients a {\it constant} value of $J_c$ in each bond gives no contribution to the exchange coupling. In the absence of strain the lattice sum allows only the isotropic $xy$ coupling.

\smallskip
Finally,  the  values of the exchange constants $(J_\| - J_\perp)$ and $J_c$ (both scalars) can vary in each link depending on the strain, and presumably one can linearize them in the manner
\begin{eqnarray}
J_{\|,ij} -  J_{\perp,ij} = \Delta J_0 + \Delta J_1 \left(\hat {\mathbf{d}}_{ij} \cdot {\boldsymbol{\varepsilon}} \cdot \hat {\mathbf{d}}_{ij} \right) \nonumber\\
J_{c,ij} =  J_c + \Delta J_{1,c} \left(\hat {\mathbf{d}}_{ij} \cdot {\boldsymbol{\varepsilon}} \cdot \hat {\mathbf{d}}_{ij} \right) \nonumber
\end{eqnarray}

\noindent where the subscript (1) denotes the coefficient of the linear term in the expansion in powers of the strain coupling.
Then by combining all these expressions, we arrive at a result for longitudinal and transverse strain couplings we had previously
\begin{eqnarray}
\Delta J_1 \,  \left[ \hat {\mathbf{d}}_{ij} \cdot {\boldsymbol{\varepsilon}} \cdot \hat {\mathbf{d}}_{ij} \right] \left( {\mathbf{s}}_i \cdot \hat {\mathbf{d}}_{ij} \right) \left( {\mathbf{s}}_j \cdot \hat {\mathbf{d}}_{ij} \right)  \nonumber
\end{eqnarray}
(Note that in this expression it is not necessary to explicitly remove the isotropic piece from the outer product because the summand includes only the traceless part of the strain tensor. In this case the lattice sum containing the $\Delta J_1$ factor does not affect isotropic spin coupling. If one were to restore the dilational strain the isotropic piece would then renormalize the effective isotropic $J$ in the $xy$ model). Rewriting this in terms of the macroscopic spin fields
\begin{eqnarray}
J \, {\mathbf{S}}_\alpha \cdot {\mathbf{S}}_\beta +  \, \Delta J_1 \, {\mathbf{S}}_\alpha \cdot \left( \sum_{{\rm nn}: \, {ij}} \, \left[ \hat {\mathbf{d}}_{ij} \cdot {\boldsymbol{\varepsilon}} \cdot \hat {\mathbf{d}}_{ij} \right]  \, \hat {\mathbf{d}}_{ij} \otimes \hat {\mathbf{d}}_{ij} \,  \right) \cdot  {\mathbf{S}}_\beta \nonumber
\end{eqnarray}
This last term has the requisite symmetries: it is even under ${\mathbf{d}} \mapsto - {\mathbf{d}}$ and $\propto {\boldsymbol{\varepsilon}}$ and is ${\cal T}$-even.

\smallskip
In the N$\rm{\acute{e}}$el state  the moments on the two sublattices are collinear (antiparallel): ${\mathbf{S}}_\alpha = -{\mathbf{S}}_\beta = \mathbb{S}$, so one can rewrite this
\begin{eqnarray}
- J \, \left| {\mathbb{S}} \right|^2 -  \, \Delta J_1 \, {\mathbb{S}} \cdot \left( \sum_{{\rm nn}: \, {ij}} \, \left[ \hat {\mathbf{d}}_{ij} \cdot {\boldsymbol{\varepsilon}} \cdot \hat {\mathbf{d}}_{ij} \right]  \, \hat {\mathbf{d}}_{ij} \otimes \hat {\mathbf{d}}_{ij} \,  \right) \cdot  {\mathbb{S}}.
\end{eqnarray}
In this expression ${\boldsymbol{\varepsilon}}$ is the externally imposed (traceless) strain and the sum is over a triad of nearest neighbor bonds away from an $\alpha$ sublattice site.

\smallskip
A similar analysis for the cross term allows for its variation as a function of strain, i.e.  using the linearize piece $J_{c,1} \left( \hat {\mathbf{d}}_{ij} \cdot {\boldsymbol{\varepsilon}} \cdot \hat {\mathbf{d}}_{ij} \right)$ we get
\begin{eqnarray}
- \Delta J_{c,1} \,  \, {\mathbb{S}} \cdot \left( \sum_{{\rm nn}: \, {ij}} \, \left[ \hat {\mathbf{d}}_{ij} \cdot {\boldsymbol{\varepsilon}} \cdot \hat {\mathbf{d}}_{ij} \right]  \, \left[ \hat {\mathbf{d_{ij}}} \otimes (\hat z \times \hat {\mathbf{d}}_{ij}) + (\hat z \times \hat {\mathbf{d}}_{ij}) \otimes \hat {\mathbf{d}}_{ij} \right] \right)\cdot  {\mathbb{S}}.
\end{eqnarray}
Summing equations 2 and 3 over a triad of bonds from an $\alpha$ site gives a $2 \times 2$ matrix parameterized by $\Delta J_1$ and $\Delta J_{c,1}$ whose principal axes denote the best and worst orientations of the N$\rm{\acute{e}}$el field in the presence of a strain $\varepsilon$. Summing the second bracket from both equations gives a form
\begin{eqnarray}
\hat {\mathbf{d}}_{ij} \otimes \left( \Delta J_1 \hat  {\mathbf{d}}_{ij}  + \Delta J_{c,1} (\hat z \times \hat {\mathbf{d}}_{ij}) \right) + \left( \Delta J_1 \hat  {\mathbf{d}}_{ij}  + \Delta J_{c,1} (\hat z \times \hat {\mathbf{d}}_{ij}) \right) \otimes \hat {\mathbf{d}}_{ij} \nonumber
\end{eqnarray}
The terms proportional to $\Delta J_1$ try to align the N$\rm{\acute{e}}$el field along the strain axes, whereas the cross product terms tip the N$\rm{\acute{e}}$el with respect to $\varepsilon$.

\smallskip
This can be seen most simply by carrying out the sum in a chiral basis where $S_\pm = S_x \pm i S_y$. If the principle axis of the strain tensor is aligned at angle $\theta_o$ one finds that the lattice sum gives an anisotropic spin coupling expressed as a matrix
\begin{gather}
\left(
  \begin{array}{cc}
    S_- \, ,& S_+ \\
  \end{array}
\right) \left(
          \begin{array}{cc}
            0 & (\Delta J_1 + i \Delta J_{c,1})e^{2 i \theta_o} \\
            (\Delta J_1 - i \Delta J_{c,1})e^{-2 i \theta_o} & 0 \\
          \end{array}
        \right)
        \left(
          \begin{array}{c}
            S_+ \\
            S_- \\
          \end{array}
        \right)  \nonumber\\
        = |\Delta J | \left[ (S_x-iS_y)^2 e^{i (2 \theta_o + \alpha) } + (S_x+iS_y)^2 e^{ - i (2 \theta_o + \alpha) } \right] \nonumber\\
        =  2|\Delta J ||S|^2 \cos( 2 (\theta -  \theta_o - \alpha/2)) \nonumber
\end{gather}
where $|\Delta J| = \sqrt{\Delta J_1^2 + \Delta J_{c,1}^2}$ and $\alpha = \arctan{[J_{c,1}/\Delta J_1]}$ defines a misalignment angle of the spin orientation $\theta$.   This shows that the role of the strain coupling is to source two units of angular momentum in the spin Hamiltonian  and that if we ignore the cross-coupling term the N$\rm{\acute{e}}$el vector would have a favored alignment along a principal axis of the strain tensor. The cross-coupling term is symmetry allowed and it misaligns the N$\rm{\acute{e}}$el field. The amount of this misalignment, which determines the direction of the N$\rm{\acute{e}}$el phase just below the phase transition is measure of the relative strength of the cross term. This might explain the experimental impression that the misalignment is  small. This would mean that the exchange is mostly determined by the separate longitudinal and transverse terms. Intuitively this makes sense since the presence of the mixed term is subtle effect having to do with the action of full crystal symmetries on the exchange. Note that this means  that it does not appear if you were to analyze an isolated bond, but the largest exchange anisotropy can be expected from the geometry within each bond. That will tend to lock the N$\rm{\acute{e}}$el field to the strain direction.

\smallskip
So the conclusion is that a linear coupling to strain provides a director that can orient the N$\rm{\acute{e}}$el vector. Ignoring ``cross coupling" of spin polarizations in each bond one would lock the favored N$\rm{\acute{e}}$el vector along a principal strain axis. Including it defines an intrinsic misorientation in the bilinear spin Hamiltonian. The previous Landau theory finds a competition between intrinsic sixfold lattice anisotropy and the direction determined by the strain. The lattice model says that the N$\rm{\acute{e}}$el field determined by the strain need not be exactly along a principal strain axis, but it would be in the absence of the cross coupling contributions to the exchange matrix.

\end{widetext}
\end{document}